\documentclass{IEEEtran}
\usepackage[ruled,linesnumbered]{algorithm2e}
\usepackage{cite}
\usepackage{graphicx}
\usepackage{amsmath}
\usepackage{array}
\usepackage{mdwmath}
\usepackage{amssymb}
\usepackage{mdwtab}
\usepackage{stfloats}
\usepackage[tight,footnotesize]{subfigure}
\usepackage{amsmath,amsthm}
\usepackage{threeparttable}
\usepackage{xcolor}
\usepackage{url}
\usepackage{algpseudocode}
\usepackage{multicol}
\usepackage{multirow}
\usepackage{setspace}
\usepackage{bm}
\usepackage{enumitem}
\algrenewcommand{\algorithmicrequire}{\textbf{Input:}}
\algrenewcommand{\algorithmicensure}{\textbf{Output:}}

\newtheorem{remark}{\it  Remark}

\newtheorem{definition}{\it  Definition}

\newcommand{\bcal}[1]{\boldsymbol{\mathcal{#1}}}
\newcommand{\hatbf}[1]{\hat{\mathbf{#1}}}

\usepackage{hyperref}
\hypersetup{
    colorlinks=true,
    linkcolor=red,
    citecolor=red,
    urlcolor=blue,
    breaklinks=true}

\begin{document}
\title{Degrees of Freedom in 3D Linear Large-Scale Antenna Array Communications---A Spatial Bandwidth Approach}
	\author{Liqin~Ding,~\IEEEmembership{Member,~IEEE,}
		Erik~G.~Str\"om,~\IEEEmembership{Fellow,~IEEE}
		and~Jiliang~Zhang,~\IEEEmembership{Senior Member,~IEEE}
		\thanks{This work was supported by H2020-MSCA-IF Grant 887732 (VoiiComm).}
		\thanks{L. Ding and E. G. Str\"om are with Department of Electrical Engineering, Chalmers University of Technology, Gothenburg, Sweden.
		}
		\thanks{J. Zhang is with Department of Electronic and Electrical Engineering,The University of Sheffield, Sheffield, S1 4ET, UK.}
	}
	\markboth{ } 
	{Ding \MakeLowercase{\textit{et al.}}: Degrees of Freedom in 3D Linear Large-Scale Antenna Array Communications---A Spatial Bandwidth Approach}
	
\maketitle
	
\begin{abstract}
	For wireless communications using linear large-scale antenna arrays, we define a receiving coordinate system and parameterization strategy  to facilitate the study of the impact of three-dimensional position and rotation of the arrays on the achievable spatial degrees of freedom (DoF) in line-of-sight (LOS) channels. An analytical framework based on spatial bandwidth analysis is developed, under which three elementary problems corresponding to three basic orthogonal receiving directions are investigated. For each of them, accurate, simple, and interpretable closed-form approximations for the achievable spatial DoF are derived, and the spatial region where a sufficient amount of spatial DoF is expected available is determined. The expressions can easily be integrated into large-scale system-level simulations. Some interesting and surprising observations are made from simulation studies based on the analytical results. For instance, the spatial bandwidth is shown to be approximately constant in almost the entire spatial multiplexing region. Moreover, in significant parts of this region, the optimal receive array orientation is not parallel with the transmitting array. 
\end{abstract}
	
	
\IEEEpeerreviewmaketitle 
	
\section{Introduction}
	
In wireless communications, information is transferred using electromagnetic waves, which propagate from transmitter to  receiver through time and space. When antenna arrays are employed, electromagnetic waves can be synthesized by manipulating multiple spatially separated radiating sources at the transmitter and measured at multiple positions at the receiver at the same instant, thereby further expanding the signal space to the spatial domain. 	In \cite{poon2005degrees}, the signal space view was adopted to study the theoretical foundation of spatial degrees of freedom (DoF) in multiple-input multiple-output (MIMO) channels between small-sized arrays in a rich scattering environment. Scatterers were treated as the sources of plane wave components that constitute the measured electric field. When their subtended solid angle from the antenna array is sufficiently large to produce a wide enough bandwidth in the spatial frequency domain (termed the wavevector domain in \cite{poon2005degrees}), spatial DoF becomes available in MIMO channels to support spatial multiplexing. The results are summarized in \cite[Chapter 7]{tse2005fundamentals}, which laid the foundation for the vigorous development of MIMO technology afterwards. A rigorous treatment of this problem using a functional analysis framework can actually be traced back to  \cite{bucci1987spatial, bucci1989degrees}. Follow-up studies with different problem settings and analysis approaches can be found in  \cite{bucci1998representation, migliore2006role, franceschetti2009capacity, franceschetti2011degrees, miller2000communicating, hanlen2006wireless}, with a recent tutorial paper \cite{miller2019waves} and a book \cite{franceschetti2018wave} providing comprehensive coverage.
	
After MIMO technology using small-sized arrays became a core element of many wireless communication standards, massive MIMO \cite{rusek2012scaling, larsson2014massive} has been adopted as a key technology for 5G New Radio (NR) in both sub-6 GHz and millimeter-wave ($24.25$--$52.6$ GHz) bands, featuring arrays consisting of on the order of $10$ antennas. There are two clear trends in array-based communications today: the trend towards even higher frequency bands such as the terahertz band ($0.1$-$10$ THz) \cite{akyildiz2018combating}; and the trend towards large-scale antenna arrays (LSAAs), characterized not only by the number of antennas of much higher order, but also by the extremely large absolute size \cite{ bjornson2019massive, de2020non, lu2021communicating}. At higher frequencies, non-line-of-sight (NLOS) communications become difficult, or even impossible, due to higher penetration losses, making LOS conditions more important. Novel LSAA-related technologies are also emerging, such as reconfigurable intelligent surfaces (RISs) \cite{basar2019wireless, di2020smart, di2020reconfigurable} and holographic MIMO  \cite{huang2020holographic, sanguinetti2021wavenumber}. 

In the meantime, to realize the vision of ubiquitous and high-capacity global coverage, non-terrestrial communications are envisioned to be key components of 6G wireless networks \cite{wu2021comprehensive, cao2018airborne, you2021towards, giordani2020non}. Satellites, high-altitude platforms (HAPs) such as manned aircrafts, and low-altitude platforms (LAPs) such as unmanned aerial vehicles (UAVs), will complement the ground infrastructure to form a three-dimensional (3D) network architecture \cite{mozaffari2018beyond}. By using collocated antenna arrays or by forming distributed MIMO through collaboration, array-based technologies are again expected to play an important role, in both air-to-ground and air-to-air connections, serving user terminals directly or providing  wireless backhaul/fronthaul \cite{ zhang2020prospective}. However, two unique characteristics are encountered: the LOS probability of the wireless channel is high due to the novel architecture; both sites and terminals are subject to 3D mobility and rotation \cite{ma2020impact, su2013maximum}. 

The lack of scattering can reduce the spatial multiplexing capability, and therefore, the LOS propagation environment is often considered unfavorable for MIMO. Fortunately, this disadvantage can be compensated using LSAAs: increasing the array aperture can create spatial DoF even under ``free-space'' propagation conditions \cite{bohagen2005construction, bohagen2007design, bohagen2007optimal}. Using a large source array, sufficiently rich spatial frequency components can be created in the electric field measured in enough long spatial duration by a large receiving array. As a consequence,  the geometric details of the arrays, namely, the shape, size, relative position (distance and direction), and rotation, become the main influencing factors on the achievable spatial DoF in the LOS MIMO channel. This makes the second characteristic, 3D mobility and rotation, exceedingly challenging.

However, how and to what extent these coupling geometric details have an impact are largely unknown. The general formula for the \textit{achievable spatial DoF} (e.g., \cite[Eq. (8.78)]{franceschetti2018wave}) 
does not directly lead to clear insights, and explicit expressions related to geometric parameters are available only in limited settings \cite{miller2000communicating, dardari2020communicating}. Filling the knowledge gaps is critical to preventing overly optimistic performance expectations or inefficient use of spatial resources. 
In this paper, we take a step forward, focusing on linear LSAAs, and propose an analytical framework to study the impact of the geometric details of the source-receiving array assembly on the achievable spatial DoF in the LOS channel between them. We also demonstrate the potential benefits that a comprehensive understanding achieved through the study can bring.

\begin{itemize}
	\item To this end, a receiving coordinate system and parameterization strategy is defined to study the problem in three basic orthogonal directions. On this basis, an analytical framework based on spatial bandwidth examination is developed, which enables us to evaluate the contribution of each receiving direction to the achievable spatial DoF under any shape, size, orientation and position conditions, and to determine the boundaries of the spatial regions with sufficient achievable spatial spatial DoF to support spatial multiplexing.   
	\item Under this unified analytical framework, closed-form approximations for the achievable spatial DoF for all three basic receiving directions are derived.  Their usefulness are two-fold. Firstly, they bring better interpretability, which can lead to deeper insights on how the achievable spatial DoF behave under different conditions and the reasons underneath. Secondly, the expressions are useful to speed up large system-level simulations that require the computation of the achievable spatial DoF.
	\item  Through numerical study, the accuracy of the approximate formulas is verified, and some surprising results related to the influence of array orientation and position are obtained. Moreover, case studies of two simple communication scenarios show that some simple rotation control solutions resulted from this study can significantly improve and lead to consistent achievable spatial DoF performance in the desired coverage areas, and thereby demonstrate the benefits the proposed analytical framework can bring. 
\end{itemize} 
	
We notice that the extensive prior knowledge behind the theoretical basis of the achievable spatial DoF, from the physical limitations of electromagnetism \cite{bucci1987spatial, bucci1989degrees, bucci1998representation, miller2000communicating} to functional analysis of wave functions  \cite{slepian1961prolate, landau1961prolate, landau1962prolate, slepian1964prolate}, may be hard to access for researchers in the fields of communication and signal processing. Therefore, with the hope that this paper can serve as a self-contained semitutorial, we revisit the problem formulation starting with the Green's function approach, restate the signal space view by clarifying the concepts of spatial frequency and spatial bandwidth, and reinterpret the achievable spatial DoF results by emphasizing the connection to array geometry. 

We remark that the amount of achievable spatial DoF (i.e., the \textit{K number}) calculated following either the general formula or the derived closed-form approximations will approach the actual achievable spatial DoF as the array size becomes sufficiently large.  So, more strictly speaking, it is better to call it the \textit{asymptotically achievable spatial DoF}. Moreover, this amount of spatial DoF can only be achieved if the electric current density distributed over the source array geometry can be manipulated as one wish and the electric field over the receiving array can be measured ideally. These conditions can be affected by many issues encountered in real system implementations (e.g., antenna radiation patterns and hardware impairments). The K number is a guide to the amount of spatial DoF  that a good array design can possibly achieve, just as the Shannon channel capacity is for the achievable rate of channel codes. However, providing solutions for real systems is beyond the scope of this paper. Furthermore, the capacity of a channel is determined by the available spatial DoF and its power gain, radiating power at the transmitter, and noise level at the receiver, together, and that the spatial DoF is a good performance indicator only at the high signal-to-noise (SNR) region. Therefore, it is also essential to study the influence of the geometry of the array assembly on the LOS channel power gain. This is also beyond the scope of this paper, but efforts are being made in this direction  \cite{dardari2020communicating, bjornson2020power, lu2021communicating}. 
	
\textit{Notations:} We use calligraphic letters for geometric entities in 3D space (e.g., volume, surface, curve, etc.), boldface calligraphic letters for electromagnetic entities, and boldface uppercase and lowercase letters for matrices and vectors respectively; $j = \sqrt{-1}$.  Given a 3D Cartesian coordinate system with coordinates  $(\mathrm x, \mathrm y, \mathrm z)$,  $\hat{\mathbf{e}}_x$, $\hat{\mathbf{e}}_y$, and $\hat{\mathbf{e}}_z$ stand for the three orthogonal standard unit basis vectors. Following the convention of electromagnetics, {row vector} is the default vector form, and  a vector in space is represented by default as $\mathbf{a} = (a_x, a_y, a_z)  = a_x \hat{\mathbf{e}}_x + a_y \hat{\mathbf{e}}_y + a_z \hat{\mathbf{e}}_z$.  $(\cdot)^\mathrm{T}$ is the transpose operation. Given two vectors $\mathbf{a}$ and $\mathbf{b}$, $\langle \mathbf{a}, \mathbf{b} \rangle = \mathbf{a} \mathbf{b}^\mathrm{T}$ is the inner product; $| \mathbf a | = \sqrt{\langle \mathbf a, \mathbf a \rangle}$ is the norm of $\mathbf{a}$, and $\hat{ \mathbf{a} } = \frac{\mathbf a}{| \mathbf a |}$ stands for its unit directional vector if $| \mathbf a | \neq 0$.

\section{Spatial DoF in LOS Channels between LSAAs}
\label{sec:2}
	
In this section, we review some existing knowledge related to the achievable spatial DoF of the LOS channels in LSAA-based communications. In particular, we clarify the concepts of spatial frequency and spatial bandwidth that will be used throughout the paper, and heuristically explain how the spatial DoF is created by the scaling of the antenna arrays. Interested readers can refer to \cite{poon2005degrees, bucci1987spatial, bucci1989degrees, bucci1998representation, miller2000communicating, hanlen2006wireless, migliore2006role, franceschetti2009capacity, franceschetti2011degrees, miller2019waves, franceschetti2018wave} for more details behind the problem formulation and the derivation of the cited results. 

\subsection{Formulating LOS Channel through Green's Function} 
\label{sec:2a}
	
Consider the electromagnetic field generated by a continuous array $\mathcal A_s$, which is composed of an infinite number of point sources of electric current distributed on an arbitrary geometric shape (e.g., volume, surface, or curve) in 3D free space.  One can think of a point source as an infinitesimal antenna composed of three Hertzian dipoles in three orthogonal directions. Arbitrarily polarized electric fields can be generated with such sources. Denoting the time-harmonic (with $\exp(j\omega t)$ convention\footnote{Monochromatic waves at (angular) frequency $\omega$ is assumed throughout this paper.}) current density at the point $\mathbf s \in \mathcal A_s$ by $\bcal{J}(\mathbf s)$, the electric field at a generic position $\mathbf{p}$ outside of $\mathcal A_s$ is given by 
\begin{equation}\label{eq:electric_field_general}
		\bcal{E}(\mathbf{p}) = \int_{\mathcal{A}_s}   \bcal{G}(\mathbf{p}, \mathbf{s}) \bcal{J}(\mathbf{s}) \, \mathrm{d}\mathbf{s} ,
\end{equation}
where $\bcal{G}(\mathbf{p}, \mathbf{s})$ stands for the dyadic Green's function, that is, the spatial impulse response (vector potential)  at the position $\mathbf p$ due to the current source at the position $\mathbf s$. 
Denoting the permittivity and permeability of free space by $\epsilon$ and $\mu$, respectively, and letting $\mathbf r = \mathbf p - \mathbf s$ be the separation vector, $r = |\mathbf r|$, and $\hatbf{r} = \mathbf r/r$, the dyadic Green's function can be explicitly expressed as follows \cite[Appendix~I]{poon2005degrees}: 
\begin{align}
	\bcal{G}(\mathbf{p}, \mathbf{s}) = & \frac{j\omega\mu}{4 \pi r } \exp(jk_0 r)  \nonumber \\
		& \Big[  \underbrace{\left(  \mathbf{I}  - \hatbf{r}^{\mathrm{T}} \hatbf{r}  \right)}_{\text{``propagating''}}
		+\underbrace{ \Big( \frac{j}{k_0 r} -\frac{1}{k_0^2 r^2} \Big)   \left( \mathbf{I} - 3 \hatbf{r}^{\mathrm{T}} \hatbf{r} \right) }_{\text{``non-propagating''}}
		\Big], 
		\label{eq:dyadic_Green_explict}
\end{align}
where $k_0 = \omega \sqrt{\epsilon \mu} = 2\pi/\lambda$ and $\lambda$ is the wavelength, and $\mathbf{I}$ stands for the $3\times 3$ identity matrix. 
	
As \eqref{eq:dyadic_Green_explict} shows, $\bcal{G}(\mathbf{p}, \mathbf{s})$ contains  a ``propagating'' component whose power decays as $r^{-2}$, as well as two ``non-propagating'' components whose power decay as $r^{-4}$ and $r^{-6}$ respectively. Only the first component contributes to the far-field radiation, since the power of the other two diminishes to a negligible amount in a few wavelengths away from the source. This \textit{a-few-wavelength-apart} condition holds for most practical wireless communication systems, as are the LSAA-based communication systems considered in this paper. This allows us to focus only on the far-field Green's function:  
\begin{align} \label{eq:dyadic_Green_farfield} 
		\bcal{G}_{\mathrm F}(\mathbf{p}, \mathbf{s}) =  
		\frac{j\omega\mu}{4 \pi r} \exp(jk_0 r) \left(  \mathbf{I}  - \hatbf{r}^{\mathrm{T}} \hatbf{r} \right) .
\end{align} 
The matrix $(\mathbf{I} - \hat{\mathbf{r}}^{\mathrm{T}} \hat{\mathbf{r}})$ restricts the oscillation direction of the radiated electromagnetic field to be perpendicular to the propagation direction. The electric field is thus given by $\bcal{E}(\mathbf{p}) = \int_{\mathcal{A}_s}  \bcal{G}_{\mathrm F}(\mathbf{p}, \mathbf{s}) \bcal{J}(\mathbf{s}) \,\mathrm{d}\mathbf{s}$, and the integrand can be regarded as the wave component at $\mathbf{p}$ generated by the point source at $\mathbf s$. Note that the dependence of $\mathbf r$, $r$ and $\hatbf{r}$ on $\mathbf s$ and $\mathbf p$ has been omitted for the sake of clarity. 
	
A continuous receiving array $\mathcal A_r$ with arbitrary geometry is employed, and the electric field $\bcal{E}(\mathbf{p})$, as a vector function of the observing position $\mathbf{p} \in \mathcal{A}_r$, is considered to be perfectly measured. It is through the mapping
\begin{equation}\label{eq:mapping}
		\mathcal{J}(\mathbf s), \; \mathbf s \in \mathcal A_s, \;  \rightarrow \;  \mathcal{E}(\mathbf{p}), \; \mathbf{p} \in \mathcal{A}_r 
\end{equation}
that the communication channel between $\mathcal A_s$ and $\mathcal A_r$ completes the task of information transmission under the free-space propagation condition.

\subsection{Spatial Frequency and Spatial Bandwidth}
\label{sec:2b}
	
\begin{figure}[!t]
	\centering
	\subfigure[ ]{\includegraphics [width = .78\linewidth, trim= 0 10 0 0, clip]{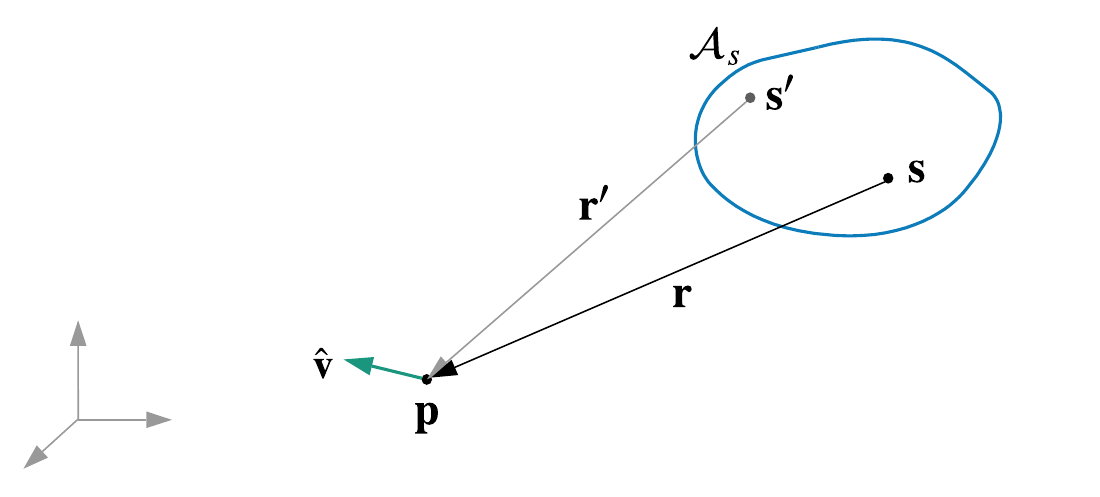}} 
	\subfigure[ ]{\includegraphics [width = .76\linewidth]{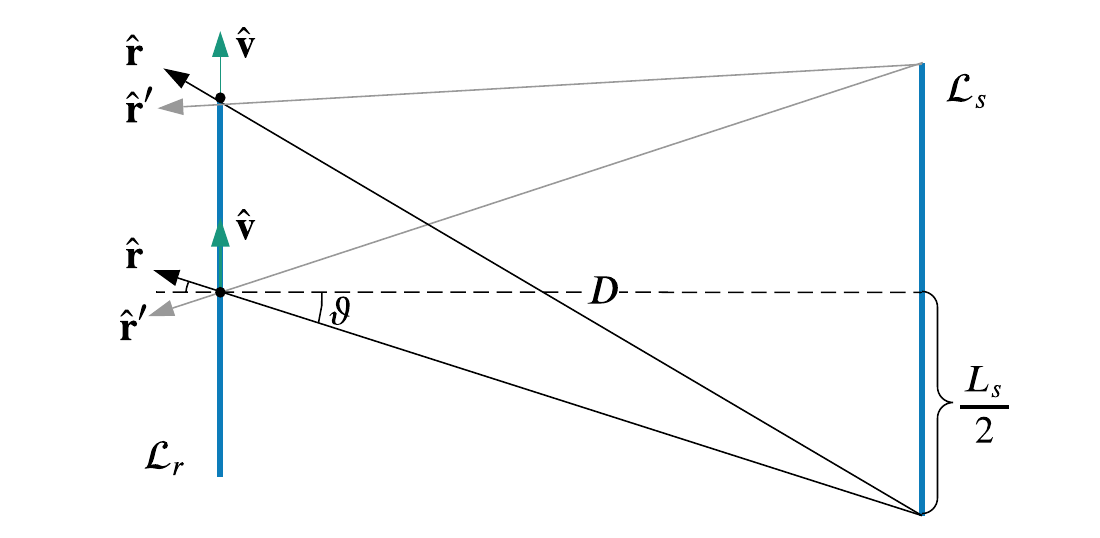}}
	\caption{Problem settings in Section II:  (a) perceiving the electric field generated by a source array $\mathcal{A}_s$ at a generic position $\mathbf p$ and in a generic direction $\hatbf{v}$; (b) the special case of two parallel linear arrays $\mathcal{L}_s$ and $\mathcal{L}_r$ in boresight direction of each other.}
	\label{fig1}
\end{figure}
	
The constant $k_0$ in the complex exponential $\exp(jk_0 r )$ in $\bcal{G}_{\mathrm F}(\mathbf{p}, \mathbf{s})$ is called the \textit{(angular) spatial frequency} or the \textit{wave number}, measured in radians per unit propagation distance. When the observing position $\mathbf{p}$ moves in a direction different from $\hatbf{r}$, however, the change in $r$ is not equal to the distance $\mathbf{p}$ moves. 
Specifically, as shown in Fig. \ref{fig1}(a), assuming that $\mathbf{p}$ moves a small step $\mathrm dl$ in the direction of $\hatbf v$, such that the incident directions of all source points on $\mathcal A_s$ can be regarded unchanged, and denoting the directional derivative of $r(\mathbf p,\mathbf s)$ in direction $\hatbf v$  by $\mathrm D_{\hatbf{v}} r(\mathbf p,\mathbf s)$, the phase of the wave component generated by $\mathbf s$, i.e., $\bcal{G}_{\mathrm F}(\mathbf{p}, \mathbf{s}) \bcal{J}(\mathbf s)\, \mathrm{d}\mathbf{s}$, is changed by $k_0 \mathrm D_{\hatbf{v}} r(\mathbf p,\mathbf s) \, \mathrm d l$. 
Letting $\kappa_{\hatbf{v}}(\mathbf p, \mathbf s)$ equal to $\frac{1}{\lambda}  \mathrm D_{\hatbf{v}} r(\mathbf p,\mathbf s)$, it is easily obtained following the definition that 
\begin{equation}\label{eq:spatial_frequency}
		\kappa_{\hatbf{v}}(\mathbf p, \mathbf s)  \triangleq  \frac{1}{\lambda}\, \mathrm D_{\hatbf{v}} r(\mathbf p,\mathbf s)
		=  \frac{1}{\lambda} \, \langle \hatbf{r}(\mathbf p,\mathbf s) , \hatbf{v} \rangle, 
\end{equation} 
because the gradient of $r(\mathbf p,\mathbf s) = |\mathbf p - \mathbf s|$ with respect to $\mathbf p$ is simply $\hatbf{r}(\mathbf p,\mathbf s)$.  The unit of $\kappa_{\hatbf{v}}(\mathbf p, \mathbf s)$ is cycle per meter. 	Therefore, $\kappa_{\hatbf{v}}(\mathbf p, \mathbf s)$ is termed the \textit{spatial frequency} of $\bcal{G}_{\mathrm F}(\mathbf{p}, \mathbf{s}) \bcal{J}(\mathbf s)\, \mathrm{d}\mathbf{s}$ in this paper.

\begin{definition}[Spatial Frequency]\label{Def:spatial_frequency}
	The spatial frequency of the wave component generated by the source point $\mathbf s$, measured at $\mathbf p$ as it moves in the  direction of $\hatbf{v}$, is given by $\kappa_{\hatbf{v}}(\mathbf p, \mathbf s) =  \frac{1}{\lambda} \langle \hatbf{r}(\mathbf p,\mathbf s) , \hatbf{v} \rangle$. 
\end{definition}

It is also clear that the spatial frequency of the wave component generated by another source point $\mathbf s'$ is different, as long as $\mathbf s'$ is not in the same incident direction when viewed from $\mathbf{p}$. In addition, as $\mathbf p$ moves in the direction of $\hatbf{v}$, the spatial frequency of all wave components constituting the measured electric field falls within the following range: 
	\begin{align*}
		\left[  
		\frac{1}{\lambda}\, \min_{\mathbf s \in \mathcal A_s } \langle \hatbf{r}(\mathbf p,\mathbf s), \hatbf{v} \rangle,  \; 
		\frac{1}{\lambda}\, \max_{\mathbf s \in \mathcal A_s} \langle \hatbf{r}(\mathbf p,\mathbf s), \hatbf{v} \rangle \right] .
	\end{align*} 
Since the inner product of two unit vectors is always between $-1$ and $1$, the above range is always a subset of $[ -\frac{1}{\lambda}, \frac{1}{\lambda}]$. Accordingly, we define the term \textit{local spatial bandwidth} used in this paper as follows.

\begin{definition}[Local Spatial Bandwidth]\label{Def:spatial_bandwidth}
	The spatial bandwidth of the electric field radiated by $\mathcal A_s$, measured locally at $\mathbf p$ as it moved in the direction of $\hatbf{v}$, is given by 
	\begin{equation} \label{eq:spatial_bandwidth}
		w_{\hatbf{v}}(\mathbf p,\mathcal A_s) \triangleq \frac{1}{\lambda} \left( \max_{\mathbf s \in \mathcal A_s} \langle \hatbf{r}(\mathbf p,\mathbf s), \hatbf{v} \rangle  
		- \min_{\mathbf s \in \mathcal A_s} \langle \hatbf{r}(\mathbf p,\mathbf s), \hatbf{v} \rangle  \right).
	\end{equation}
\end{definition}
	
Therefore, only when the size of $\mathcal A_s$ seen from $\mathbf p$ is large enough to cause a significant spread in the incident direction $\hatbf{r}$, can it possibly create a non-zero spatial bandwidth in the electric field measured at $\mathbf p$. Otherwise, if the size of $\mathcal A_s$ is too small or the propagation distance is too long, there is practically no difference in $\hatbf{r}$, $\bcal{E}(\mathbf{p})$ has only one spatial frequency component and the spatial bandwidth is zero.
	
\begin{remark}
We note that in \cite{bucci1987spatial, bucci1989degrees, bucci1998representation, migliore2006role, franceschetti2009capacity, franceschetti2011degrees}, spatial frequency and spatial bandwidth actually refer to $2\pi \kappa_{\hatbf{v}}(\mathbf p, \mathbf s)$ and $2\pi w_{\hatbf{v}}(\mathbf p,\mathcal A_s)$, and therefore, share the same unit with $k_0$. We chose Definition \ref{Def:spatial_frequency} and \ref{Def:spatial_bandwidth} to better make the analogy with the frequency and bandwidth of a time signal $\exp(j 2\pi f t)$, which will ease the discussion. 
\end{remark}
	
It should be noted that as long as $\mathcal A_s$ is not absolutely symmetric in all three dimensions, its shape and size seen from $\mathbf p$ is dependent on the relative position of $\mathbf p$. Moreover, the spatial bandwidth is affected by the choice of $\hatbf{v}$, which is restricted by the geometry of the receiving array $\mathcal A_r$. When $\mathcal A_r$ is an one-dimensional (1D) smooth curve $\mathcal C_r$ or line segment $\mathcal L_r$, the observing position can be parameterized by a single parameter $l$ as $\mathbf{p}(l)$, $0\leq l \leq L_r$, where $L_r$ is the total arc length of the array. In this case, $\bcal{E}(\mathbf{p})$ is a 1D spatial signal and can be written as $\bcal{E}(l)$. Given a curve array $\mathcal C_r$, the direction in which the observing position moves is simply given by the unit tangent vector to the curve, denoted by $\hatbf{v}(l)$, $0\leq l \leq L_r$. In the special case of a linear array $\mathcal L_r$, $\hatbf{v}(l)$ becomes a constant. Even so, as $L_r$ increases, one will observe the gradual change in the spatial frequency components and spatial bandwidth of $\bcal{E}(l)$ with $l$, because $\hatbf{r}(\mathbf p(l), \mathbf s)$ changes with $l$.

\subsection{Spatial Degrees of Freedom}
\label{sec:2c}
	
Mathematically, the number of spatial DoF available in the LOS communication channel between $\mathcal A_s$ and $\mathcal A_r$ is the number of eigenmodes admitted by the mapping \eqref{eq:mapping} from the source signal space to the receiving signal space through the linear operator $\bcal{G}_{\mathrm F}$ \cite{bucci1987spatial, bucci1989degrees, bucci1998representation, franceschetti2009capacity, franceschetti2011degrees}. Each eigenmode is associated with  an input function in the source signal space and a coupling output function in the receiving signal space.  All source functions are mathematically orthogonal, as are all the  coupling receiving functions, and they constitute the basis functions of the source signal and the receiving signal, respectively. 	These functions, as well as the associated eigenvalues (i.e., coupling coefficients), can be found by solving the singular-value decomposition (SVD) problem\footnote{For an intuitive understanding, one may reflect on performing SVD on discrete MIMO channel matrices in traditional MIMO studies, where the number and quality of eigenmodes are evaluated by rank and condition number, respectively.}, which consists of a pair of coupled eigenfunction problems. In general, the number of eigenfunction pairs is infinite. However, due to the inherent band-limiting nature of the mapping \eqref{eq:mapping}, only a limited number of eigenvalues are \textit{significant}. After them is a transition zone, where the eigenvalues drop rapidly to near zero.  When the required level of representation accuracy $\epsilon$, in the energy sense, is specified, a mathematically precise definition of DoF can be given: Denoting the eigenvalues sorted in decreasing order as $\sigma_1^2 \geq \sigma_2^2 \geq \cdots \geq  \sigma_k^2 \geq \cdots$, the number of DoF, $K_\epsilon$, is given by the index such that $\sigma_{K_\epsilon -1}^2 > \epsilon$ and $\sigma_{K_\epsilon}^2 \leq \epsilon$, which ensures that any receiving function can be represented using the first $K_\epsilon$ eigenfunctions with a squared error upper limited by $\epsilon$. For a communication link, $\epsilon$ should be determined by the overall noise level.  Given a band-limited mapping as \eqref{eq:mapping}, by choosing a sufficiently large $K_\epsilon$, any desired level of representation accuracy can be achieved. It turns out that given a 1D receiving array, $K_\epsilon$ can be approximated by \cite{franceschetti2011degrees}
	\begin{equation}\label{eq:spatial_DoF}
		K  =  \int_{\mathcal{I}} w_{\hatbf{v}(l)}(\mathbf p(l), \mathcal A_s) \; \mathrm d l, 
	\end{equation} 
where $\mathcal{I}$ stands for the effective integration range (see Remark \ref{remark1} below). This value will be referred to as the \textit{K number} of the achievable spatial DoF in the rest of the paper.
	
\begin{remark}\label{remark1}
	When the source array $\mathcal A_s$ is a convex source volume $\mathcal V_s$ that expands in all three spatial dimensions, the effective integration range of a 1D receiving array of total arc length $L_r$ outside $\mathcal A_s$ is simply given by $\mathcal{I} = [0,L_r]$. 	When $\mathcal A_s$ is dimensional deficient, however, the radiated electric field in the 3D space will contain certain symmetry or periodicity\footnote{For instance, the electric fields radiated by a planar array in two half-spaces separated by the plane containing the array are mirror images of each other, and the electric field radiated by a linear array observed on a right section of any cylindrical surface centered on the array is identical.}. If $\mathcal L_r$ is randomly placed in space, the electric field observed on a part of it may be completely or partly correlated with the electric field observed on the rest part. In other words, this part of the spatial domain contributes no additional DoF, or less than the amount returned by the integration over it. Thus, determining the effective integration range for \eqref{eq:spatial_DoF} is essential to obtain the correct K number. 
\end{remark}
	
The formula corresponding to \eqref{eq:spatial_DoF} for the special setting where two linear arrays are placed in the boresight direction of each other, as shown in Fig. \ref{fig1}(b), but with much greater separation $D$, is better known. Denote their lengths by ${L}_s$ and ${L}_r$, respectively, and assume $D \gg L_s$ and $D \gg L_r$. The achievable  spatial DoF in the LOS channel is approximately given by \cite[Eq. (67)]{miller2000communicating} 
	\begin{equation}\label{eq:Kparallel_1}
		K_{\mathrm{parallel}} = \frac{L_s L_r}{\lambda D}. 
	\end{equation}
This result can be easily derived based on the spatial bandwidth analysis under two approximations: Firstly, the spatial bandwidth over the entire receiving array is approximately considered constant, denoted by $w_0$. Secondly,  adopting the small-angle approximation $\sin\vartheta \approx \tan \vartheta$, the range of spatial frequency of the measured electric field at the center of $\mathcal L_r$ is approximately given by 
	\begin{equation*}
		\Big[ \frac{\langle \hatbf{r}', \hatbf{v} \rangle}{\lambda}, \frac{\langle \hatbf{r}, \hatbf{v} \rangle }{\lambda} \Big] 
		= \Big[- \frac{\sin\vartheta}{\lambda} , \frac{\sin\vartheta}{\lambda} \Big]  
		\approx \Big[ - \frac{L_s}{2\lambda D} ,  \frac{L_s}{2\lambda D}  \Big].  
	\end{equation*}
Thus, $w_0 = \frac{L_s}{\lambda D}$. Moreover, the effective integration range for \eqref{eq:spatial_DoF} is given by $\mathcal{I} = [0,L_r]$ in this setting. 	Substituting them into \eqref{eq:spatial_DoF}, \eqref{eq:Kparallel_1} is obtained immediately. 

As an analogy, consider a time signal that is bandlimited to $[-B, B]$ Hz. If we sample the signal with the Nyquist sampling rate $2B$ Hz, then $2BT$ samples are accumulated during $T$ seconds. It turns out that, as $B$ and/or $T$ becomes sufficiently large, $2BT$ is also the smallest number of basis functions needed to represent the time signal with negligible error during $T$ seconds \cite{slepian1961prolate, landau1961prolate, landau1962prolate}. In that sense, $2BT$ is the number of DoF of the set of bandlimited signals. By rewriting \eqref{eq:Kparallel_1} as $ K_{\mathrm{parallel}} = 2 \frac{L_s}{2\lambda D} L_r$, the similarity between  $ K_{\mathrm{parallel}}$ and $2BT$ can be easily seen. As another analogy, $K$ and $K_{\mathrm{parallel}}$ approach the actual achievable spatial DoF better and better as the size of the arrays increases  \cite[Chapter 8.4]{franceschetti2018wave}. Therefore, it should be understood that, strictly speaking, the K number is the asymptotically achievable spatial DoF in the LOS channel. 
	
\subsection{Discussions and Remarks}
	
The spatial bandwidth viewpoint can be briefly summarized as follows: Regarding the measured electric field as a bandlimited spatial signal, increasing the source array size tends to increase the local spatial bandwidth, and increasing the receiving array size tends to increase the observation interval. Hence, increasing the size of the source and/or receiving array tends to increase the achievable spatial DoF, i.e., the K number.

The K number depends only on the geometry of the arrays, and not on any practical issues such as antenna radiation pattern and mutual coupling effects \cite{ chen2018review}, which greatly simplifies the evaluation of spatial DoF. However, as emphasized in the introduction, the K number stands for the achievable spatial DoF under the assumption that we have complete control over the current density distribution on the source array and perfect perception of the electric field on the receiving array, which is susceptible to many practical issues encountered in the system implementation. Their effects will be shown on the actual eigenvalues, which determine the power gain distribution among the eigenmodes, and thus eventually on the actual achievable channel capacity \cite{miller2000communicating}. 

A distinct feature of the spatial domain that adds to the complexity of the problem, is the freedom to place, rotate, and move the arrays in 3D space. Through affecting the value range of $\hatbf r$, $\hatbf v$, and sometimes $\mathcal I$, all geometric details of the array assembly, i.e., shape, size, distance, relative direction, and rotation/orientation in all spatial dimensions will affect the amount of the achievable spatial DoF in the LOS channel.
		
\begin{remark}
	When the geometry of the receiving array is given by a two-dimensional (2D) surface, the observing position $\mathbf p$ is given the freedom to move in two orthogonal  directions in space, and thus, the measured electric field can be regarded as a 2D spatial signal \cite{bucci1998representation, dardari2020communicating}. The local spatial bandwidth is determined by the area of the region spanned by the projection of $\hatbf{r}$ on the surface element centered at $\mathbf p$, and the K number is given by surface integral \cite[Eq. (28)]{dardari2020communicating}.  In theory, the receiving array can also be a 3D volume, but only the part of the surface exposed to the LOS direction of the source array contributes actually to the expansion of the spatial domain of the perceived electric field. The surface array is of strong relevance to many recent research topics including holographic MIMO and RISs.  However, we will not discuss it in detail in this paper because the nature of the spatial frequency viewpoint remains the same.
\end{remark}
	
\begin{remark}
	It can also be inferred from  \eqref{eq:spatial_bandwidth} that with the continuous scaling of the receiving array in space, the local spatial bandwidth diminishes to zero, because the spread in the incident direction $\hatbf r$ will be reduced to zero. Therefore, regardless of the array geometry, the total amount of achievable spatial DoF admitted by a bounded source array is limited \cite{franceschetti2011degrees}, which means that under a transmit power constraint, the spatial channel capacity is bounded, even as the size of the receiving array grows without bound. 
\end{remark}
	
\begin{remark}
	In the above discussions, we have deliberately avoided the fact that \eqref{eq:mapping} is a mapping between two vector functions to focus on the spatial domain. The mapping is through a rank-$2$ matrix $(\mathbf{I} - \hat{\mathbf{r}}^{\mathrm{T}} \hat{\mathbf{r}})$. This means that when polarization is fully explored, it will double the overall DoF available in the LOS channel between the two arrays \cite{poon2005degrees}, provided that they are at least a few wavelengths apart such that the far-field Green's function \eqref{eq:dyadic_Green_farfield} applies. 
\end{remark}
	
\section{Proposed Methodology}
\label{sec:4}
	
In the rest of the paper, we make an initial effort to establish an analytical framework to enable an in-depth study of the available conditions and the main influencing factors of the spatial DoF in the LOS channel, with the focus on linear arrays.  In Section \ref{sec:2}, we heuristically show that all geometric details of the source-receiving array assembly matter.  However, the general integral formula \eqref{eq:spatial_DoF} does not provide much indication of how and to what extent those coupling geometric factors affect the K number. In fact, the impact of the distance factor alone is already nontrivial. 

As the distance between the two arrays gradually decreases from very large to very small, we expect the LOS channel to be in one of three regions: \textit{i)} The source array generates a single frequency component in the measured electric field, no spatial bandwidth is created, and the LOS channel has no spatial multiplexing capability. \textit{ii)} A spread in the incident direction occurs, creating a nonzero spatial bandwidth in the observed electric field, which remains approximately constant over the entire receiving array. Eventually, a sufficient amount of spatial DoF is accumulated to support spatial multiplexing. \textit{iii)} The spread in the incident direction seen at different positions of the receiving array shows noticeable differences, causing the spatial bandwidth to vary over the receiving array. This variation needs to be taken into account, for example, for optimal antenna placement. Therefore, it is of interest to determine the boundaries between these regions. Moreover, it can be inferred that the thresholds will also depend on the relative direction and rotation/orientation of the arrays. 
	
We aim for more insightful expressions than the general formulas \eqref{eq:spatial_bandwidth} and  \eqref{eq:spatial_DoF} to explicitly connect the geometric parameters of the array assembly to the K number.	Since the spatial relation between the arrays can no longer be described using a single separation vector, a new parameterization strategy is needed. In this section, we first propose a receiving coordinate system and an associated parameterization strategy, which lead to the study of three elementary problems involving three orthogonally orientated linear receiving arrays, and eventually, simple spatial bandwidth expressions with better interpretability. The proposed analytical tools are introduced afterwards.

\subsection{Receiving Coordinate System and Parameterization}
\label{sec:4a}

\begin{figure}[!t]
	\centering
	\includegraphics[width=.62\linewidth, trim= 0 0 0 0, clip]{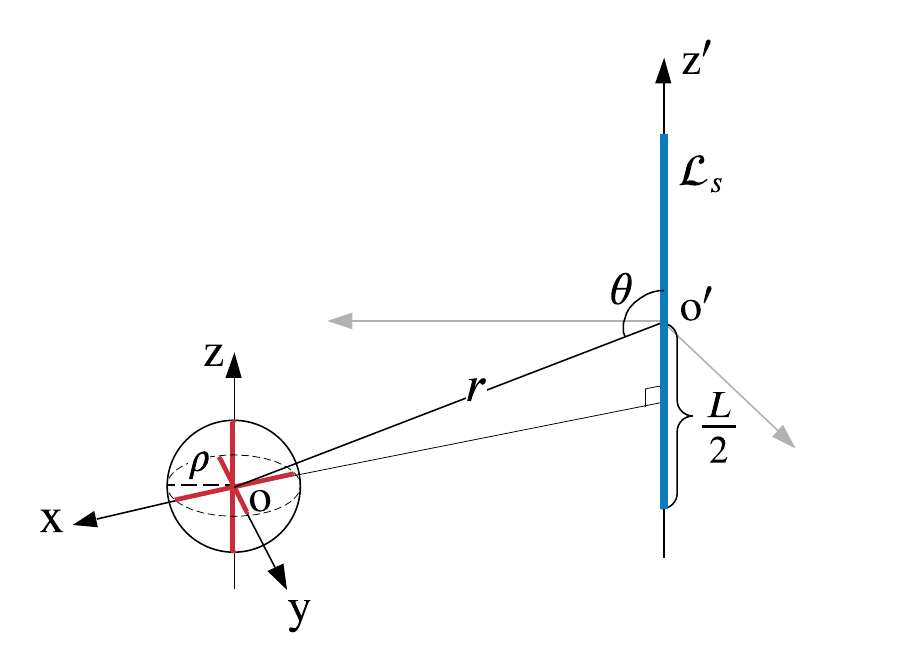} 
	\caption{The proposed local receiving coordinate system and the associated parameters. The three hypothetical linear receiving arrays are depicted by the three red bars.} 
	\label{fig2}
\end{figure}

As shown in Fig. \ref{fig2}, the global coordinate system is assumed to have the origin $\mathrm o'$ set at the center of the linear source LSAA $\mathcal L_s$ and the $\mathrm z'$-axis aligned with $\mathcal L_s$. The distance between $\mathrm o'$ and the center of the linear receiving array $\mathcal L_r$, denoted by $\mathrm o$, is given by $r$. For clarity, the lengths of $\mathcal L_s$ and $\mathcal L_r$ is denoted by $L$ and $2\rho$, respectively.  Under the ideal isotropic point source assumption, the revolving of the receiving array around the $\mathrm z'$-axis will not affect the electric field it observes. Therefore, it is sufficient to parameterize the relative direction of the receiving array placement by the polar angle $\theta$ of the $\mathrm o'$-$\mathrm o$ connecting line from the $\mathrm z'$-axis, which is regarded as the zenith direction.

To focus and facilitate the discussion on the receiving side, a right-handed local receiving coordinate system is defined with the origin set at $\mathrm{o}$ and three orthogonal directions $\{\hat{\mathbf{e}}_{x}, \hat{\mathbf{e}}_{y} , \hat{\mathbf{e}}_{z} \}$ set as follows:  $\hat{\mathbf{e}}_{z}$ is parallel to the global $\mathrm z'$-axis and to $\mathcal L_s$, $\hat{\mathbf{e}}_{x}$ lies on the plane determined by $\mathrm{o}$ and $\mathcal L_s$ and is orthogonal to $\hat{\mathbf{e}}_{z}$, and $\hat{\mathbf{e}}_{y}$ is perpendicular to this $\mathrm{o}$-$\mathcal L_s$ plane. $\mathcal L_r$ can be arbitrarily rotated, in the direction given by a unit directional vector $\hatbf v$ in the receiving coordinate system. It is often considered that the most favorable orientation is when $\mathcal L_s$ and $\mathcal L_r$ are parallel. Existing linear antenna array designs for LOS MIMO with arbitrary orientations \cite{bohagen2007design, bohagen2007optimal} actually equivalent to considering only projections in this direction. 	While the $\hat{\mathbf{e}}_{y}$ direction is often considered irrelevant and ignored in linear array studies. However, from the definition equation \eqref{eq:spatial_bandwidth} we can see that a nonzero spatial bandwidth can also be created in this direction if the two arrays are close enough. Therefore, $\hat{\mathbf{e}}_{z}$ and $\hat{\mathbf{e}}_{y}$ are selected to be two basis directions, and $\hat{\mathbf{e}}_{x}$ is chosen to satisfy the right-hand rule.

With the above definitions, the geometric details of a linear source-receiving array assembly $(\mathcal L_s, \mathcal L_r)$ can be described using the four parameters $(L, \rho, r,\theta)$ and the unit directional vector $\hatbf v$: $L$ and $2\rho$ are the array lengths, $r$ and $\theta$ describe their distance and relative direction, and $\hatbf v$ the relative orientation. Note that $L > 0$,  $r>0$, $\theta \in [0, \pi]$, and $\rho > 0$ are always assumed. For clarity in the subsequent analysis, we define two parameter vectors $\mathbf{\Omega} \triangleq (L, r, \theta)$ and $\overline{\mathbf{\Omega}} \triangleq (L, \rho, r, \theta)$. 
	
Finally, we define three hypothetical linear receiving arrays orientated in $\hat{\mathbf{e}}_{x}$, $\hat{\mathbf{e}}_{y}$, and $ \hat{\mathbf{e}}_{z} $ directions, denoted by $\mathcal L_x$, $\mathcal L_y$ and $\mathcal L_z$, respectively, as depicted using three red bars in Fig.~\ref{fig2}. Denoting their associated K number by $K_x$, $K_y$, and $K_z$, respectively, and that with $\mathcal L_r$ by $K_{\hatbf{v}}$, the following relations can be found: 
\begin{align*}
	\max (K_x, K_y, K_z) \leq \max_{\hatbf{v}} K_{\hatbf{v}}, \\
	\min (K_x, K_y, K_z) \geq \min_{\hatbf{v}} K_{\hatbf{v}}. 
\end{align*}
These relations allow us to focus the study on three elementary problems involving linear arrays only, namely, on the $ (\mathcal L_s, \mathcal L_x)$, $ (\mathcal L_s, \mathcal L_y)$, and $ (\mathcal L_s, \mathcal L_z)$  assemblies.

\subsection{Analytical Framework}
\label{sec:4b}
	
Consider the arbitrarily rotated linear receiving array $\mathcal L_r$. A point on it can be expressed as $l \hatbf{v}$, with $l\in [-\rho, \rho]$. The local spatial bandwidth of the electric field perceived at this point is given by 
\begin{equation}
	\label{eq:spatial_bandwidth_RCS}
	w_{\hatbf{v}} (l; \mathbf{\Omega}) =  \frac{1}{\lambda} \left( \max_{\mathbf s \in \mathcal L_s} \langle \hatbf{r}(l,\mathbf s), \hatbf{v} \rangle -  \min_{\mathbf s \in \mathcal L_s} \langle  \hatbf{r}(l,\mathbf s), \hatbf{v} \rangle \right),
\end{equation}
where $\hatbf{r}(l,\mathbf s)$ stands for the unit incident direction vector from the source point $\mathbf s \in \mathcal L_s$ to $l \hatbf{v}$, given also in the receiving coordinate system. 
	
We denote the effective integration range by $\mathcal I_{\hatbf{v}}(\overline{\mathbf{\Omega}} )$ and leave the discussion regarding the three hypothetical linear receiving arrays to the next section. With this notation, the achievable spatial DoF by $\mathcal L_r$ is given by 
\begin{equation}\label{eq:spatial_DoF_RCS}
	K_{\hatbf{v}}(\overline{\mathbf{\Omega}} )  =  \int_{ \mathcal I_{\hatbf{v}} (\overline{\mathbf{\Omega}} ) } w_{\hatbf{v}} (l; \mathbf{\Omega}) \, \mathrm d l .
\end{equation}
	
In addition, we define
\begin{align}
	w_{\hatbf{v}}^{\mathrm{max}} (\overline{\mathbf{\Omega}})  & \triangleq  \max_{l \in \mathcal I_{\hatbf{v}}(\overline{\mathbf{\Omega}}) } w_{\hatbf{v}} (l; \mathbf{\Omega}) , \label{eq:spatial_bandwidth_max}  \\
	w_{\hatbf{v}}^{\mathrm{min}} (\overline{\mathbf{\Omega}}) & \triangleq  \min_{l \in \mathcal I_{\hatbf{v}}(\overline{\mathbf{\Omega}}) } w_{\hatbf{v}} (l; \mathbf{\Omega}),  \label{eq:spatial_bandwidth_min}
\end{align}
and 
\begin{equation} \label{eq:spatial_bandwidth_range}
		w_{\hatbf{v}}^{\mathrm{range}} (\overline{\mathbf{\Omega}})   \triangleq 
		w_{\hatbf{v}}^{\mathrm{max}} (\overline{\mathbf{\Omega}})  - w_{\hatbf{v}}^{\mathrm{min}} (\overline{\mathbf{\Omega}}). 
\end{equation}
The dependence on $\overline{\mathbf{\Omega}}$ emphasizes the fact that the behavior of the local spatial bandwidth is determined by the geometric details of the linear array assembly. We note that  $w_{\hatbf{v}}^{\mathrm{range}}$ indicates the severity of the change in the local spatial bandwidth over $\mathcal L_r$. Under certain circumstances, the change is negligible such that constant approximations can be made.  Both $w_{\hatbf{v}}^{\mathrm{max}} (\overline{\mathbf{\Omega}})$ and $w_{\hatbf{v}}^{\mathrm{min}} (\overline{\mathbf{\Omega}})$ may be used for constant approximation, leading to an upper and an lower bound on the actual K number $K_{\hatbf{v}}(\overline{\mathbf{\Omega}} )$: 
\begin{align}
	K^{\mathrm u}_{\hatbf{v}}(\overline{\mathbf{\Omega}} ) \triangleq  w_{\hatbf{v}}^{\mathrm{max}} (\overline{\mathbf{\Omega}}) \cdot \left| \mathcal I_{\hatbf{v}} (\overline{\mathbf{\Omega}} ) \right| , \label{eq:spatial_DoF_RCS_apx} \\
	K^{\mathrm l}_{\hatbf{v}}(\overline{\mathbf{\Omega}} ) \triangleq  w_{\hatbf{v}}^{\mathrm{min}} (\overline{\mathbf{\Omega}}) \cdot \left| \mathcal I_{\hatbf{v}} (\overline{\mathbf{\Omega}} ) \right| ,
\end{align}
where $| \mathcal I_{\hatbf{v}} (\overline{\mathbf{\Omega}} ) | $ stands for the Lebesgue measure (i.e., length in the 1D case) of $\mathcal I_{\hatbf{v}} (\overline{\mathbf{\Omega}} )$. Both bounds are tight when the change of $w_{\hatbf{v}} (l; \mathbf{\Omega})$ over $ \mathcal I_{\hatbf{v}}(\overline{\mathbf{\Omega}})$ can be ignored. 
	
When the change of $w_{\hatbf{v}} (l; \mathbf{\Omega})$ is significant, a more accurate but tractable approximation of $K_{\hatbf{v}}(\overline{\mathbf{\Omega}} )$ is desired. It is not difficult to infer from the definition that the change of $w_{\hatbf{v}} (l; \mathbf{\Omega})$ is nonlinear with $l$. It can also be seen from the analysis in the next section that under different geometric conditions, the maximum and  minimum values of $w_{\hatbf{v}} (l; \mathbf{\Omega})$ can appear at any point on the receiving array. Nevertheless, we propose to approximate the local spatial bandwidth using a linear function, whose value increases from $w_{\hatbf{v}}^{\mathrm{min}} (\overline{\mathbf{\Omega}})$ to $w_{\hatbf{v}}^{\mathrm{max}} (\overline{\mathbf{\Omega}})$ within the effective integration range $\mathcal I_l(\overline{\mathbf{\Omega}} )$. Accordingly,  $K_{\hatbf{v}}(\overline{\mathbf{\Omega}} )$ can be approximated using the following linear formula: 
\begin{equation}
	{K}^{\mathrm a}_{\hatbf{v}}(\overline{\mathbf{\Omega}} ) \triangleq  
	\frac{1}{2} \left( w_{\hatbf{v}}^{\mathrm{min}} (\overline{\mathbf{\Omega}}) + w_{\hatbf{v}}^{\mathrm{max}} (\overline{\mathbf{\Omega}}) \right)
	\cdot \left| \mathcal I_{\hatbf{v}} (\overline{\mathbf{\Omega}} ) \right| \label{eq:spatial_DoF_RCS_apx2} .
\end{equation}
The goodness of this approximation will be shown using the numerical results in Section \ref{sec:6}. 
	
\subsection{Spatial-Multiplexing Region} 
\label{sec:4c}
	
To form a comprehensive understanding of the available conditions of a sufficient amount of spatial DoF in the LOS channel, we define the concept of  \textit{spatial multiplexing region} for the generic linear array assembly $ (\mathcal L_s, \mathcal L_r)$ as follows. 
	
\begin{definition}[Spatial Multiplexing Region]\label{Def:spatial_multiplexing_region}
	Given a linear array assembly $(\mathcal L_s, \mathcal L_r)$, the spatial multiplexing region, denoted by $\mathcal R_{\hatbf{v}}(\overline{\mathbf{\Omega}}, K_0)$, is the set of locations, where the achievable spatial DoF in the LOS channel between $\mathcal L_s$ and $\mathcal L_r$  reaches a given threshold $K_0$. Namely, 
	\begin{equation}
		\mathcal R_{\hatbf{v}}(\overline{\mathbf{\Omega}}, K_0)	\triangleq \left\{ \mathbf r \in \mathbb R^3: K_{\hatbf{v}}(\overline{\mathbf{\Omega}} ) \geq K_0 \right\}. 
	\end{equation} 
\end{definition}
	
In other words, $\mathcal R_{\hatbf{v}}(\overline{\mathbf{\Omega}}, K_0)$ is the spatial region surrounding $\mathcal L_s$, such that when $\mathcal L_r$ is located in it, the K number, given by \eqref{eq:spatial_DoF_RCS}, is at least $K_0$, where $K_0$ is usually a small number, but sufficient to ensure the required spatial multiplexing capability. When there is no specific requirement for the spatial multiplexing capability, $K_0 = 1$ is suggested.  The boundary of the spatial multiplexing region is given by the distance threshold at polar angle $\theta$ such that $K_{\hatbf{v}}(\overline{\mathbf{\Omega}} ) = K_0$, for all $\theta \in [0,\pi]$. 
	
\begin{remark}
	It will be verified using the numerical results in Section \ref{sec:6c} that  $K_0=1$ serves as a good mathematical indicator of the availability of spatial multiplexing capability in the LOS channel. However, it should not be forgotten that the result given by \eqref{eq:spatial_DoF_RCS} is an asymptotically good approximation of the actual number of DoF only when $w_{\hatbf{v}} (l; \mathbf{\Omega})$ and/or $| \mathcal I_{\hatbf{r}} (\overline{\mathbf{\Omega}} ) |$ is large enough.
\end{remark}
	
Conceptually, the spatial multiplexing region can be divided into two parts: a part where a constant spatial bandwidth approximation leads to negligible error in K number calculation; and a part where the  approximation error is not negligible. We name them the \textit{constant-bandwidth spatial multiplexing region} and the \textit{non-constant-bandwidth spatial multiplexing region}, respectively. 

Suppose we use $w_{\hatbf{v}}^{\mathrm{max}} (\overline{\mathbf{\Omega}}) $ as the constant spatial bandwidth approximation, and $ \Delta K$ is the maximum tolerable K number calculation error. The non-constant-bandwidth spatial multiplexing region, denoted by $\mathcal R^{\mathrm{nc}}_{\hatbf{v}}(\overline{\mathbf{\Omega}}, \Delta K)$, is then the subset of $\mathcal R_{\hatbf{v}}(\overline{\mathbf{\Omega}}, K_0)$ such that 
\begin{equation}\label{eq:CBSMR}
	K^{\mathrm u}_{\hatbf{v}}(\overline{\mathbf{\Omega}} ) - K_{\hatbf{v}}(\overline{\mathbf{\Omega}} ) > \Delta K. 
	\end{equation}
The boundary of $\mathcal R^{\mathrm{nc}}_{\hatbf{v}}(\overline{\mathbf{\Omega}}, \Delta K)$ at polar angle $\theta$ can be found by solving  the equation $K^{\mathrm{u}}_{\hatbf{v}}(\overline{\mathbf{\Omega}} ) - K_{\hatbf{v}}(\overline{\mathbf{\Omega}} ) =  \Delta K$.

By replacing the directional vector $\hatbf{v}$ with $\hat{\mathbf{e}}_{x}$, $\hat{\mathbf{e}}_{y}$, and $\hat{\mathbf{e}}_{z}$, the definitions given in Section \ref{sec:4b} and \ref{sec:4c} extend naturally to the three hypothetical linear receiving arrays $\mathcal L_x$, $\mathcal L_y$ and $\mathcal L_z$. For them, subscripts $x$, $y$, and $z$ will be used in all the related notations for the sake of simplification.
	
\section{Spatial Bandwidth Analysis in Three Receiving Directions}
\label{sec:5}
	
In this section, we derive analytical expressions for the local spatial bandwidth observed on $\mathcal L_z$,  $\mathcal L_x$, and $\mathcal L_y$ (in this order, considering the decreasing contribution to spatial DoF they are capable of making). Explicit expressions for the maximum and minimum of the local spatial bandwidth, as well as the linear approximation of the K number, will also be given. It should be emphasized that although the derived expressions can be applied to any choices of $\overline{\mathbf{\Omega}}$, the results will not reflect the correct physical reality when $r$ on the order of a few wavelengths, since then the far-field Green's function will not be applicable to the entire receiving array. 
	
\begin{figure*}[t]
	\centering
	\subfigure[$\mathcal L_z$]{\includegraphics [height = 4.8cm, trim= 0 10 0 0, clip]{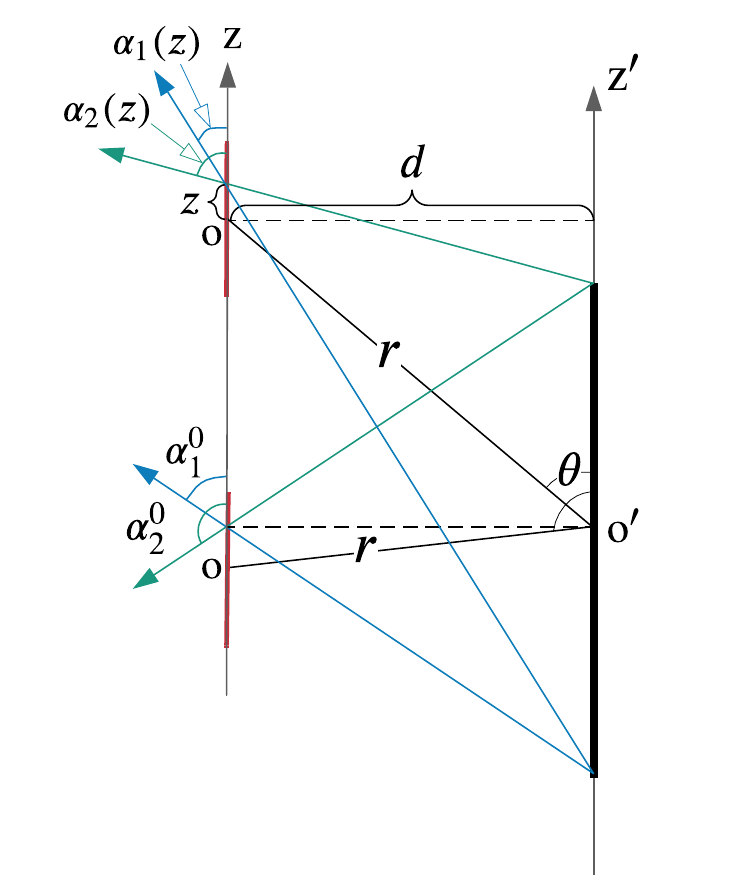}} \hspace{1.5em}
	\subfigure[$\mathcal L_x$]{\includegraphics [height = 4.8cm, trim= 0 10 0 0, clip]{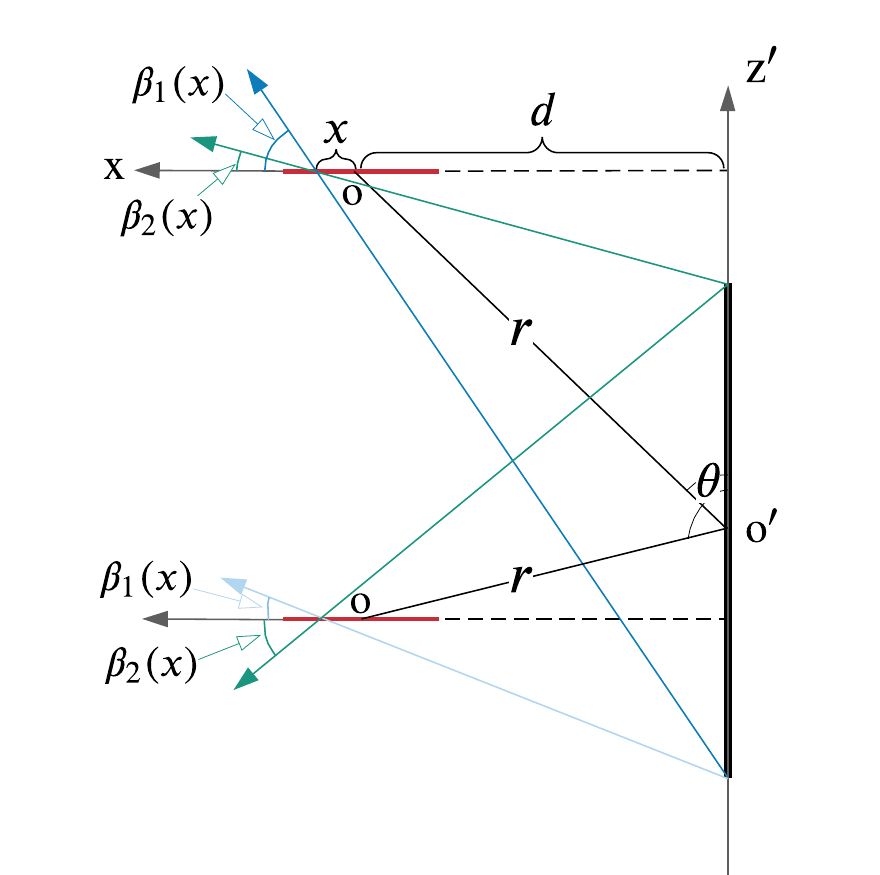}}\hspace{1.5em}
	\subfigure[$\mathcal L_y$]{\includegraphics [height = 4.8cm, trim= 0 10 0 0, clip]{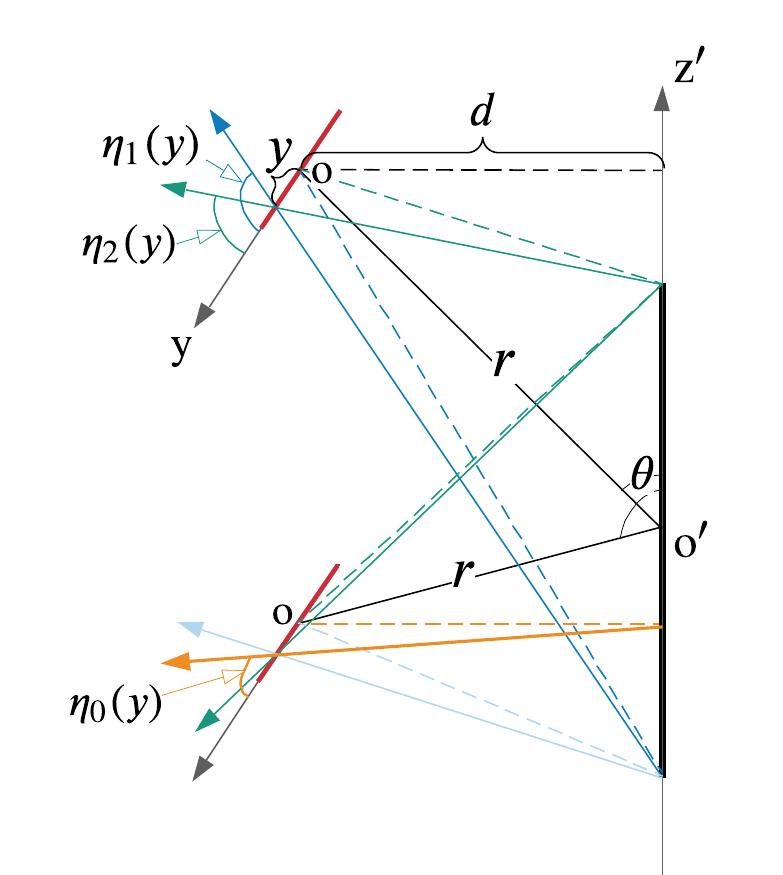}}
	\caption{Illustration of the coordinates and key angles used for local spatial bandwidth analysis for the three hypothetical linear receiving arrays.} 
	\label{fig3}
\end{figure*}
	
\subsection{In \texorpdfstring{$ \hat{\mathbf{e}}_{z} $}{ez} Direction}
\label{sec:5a}
	
Denote $w_{z}(z;  \mathbf{\Omega})$ to be the local spatial bandwidth of the electric field perceived at a point on $\mathcal L_z$ specified by coordinate $z$, which is depicted in Fig. \ref{fig3} (a). Since there is no symmetry in the radiated electric field in $ \hat{\mathbf{e}}_{z} $ direction, the effective  integration range for  $w_{z}(z;  \mathbf{\Omega})$ is given by $\mathcal I_z(\overline{\mathbf{\Omega}} ) = [-\rho, \rho]$. 
	
As shown in Fig. \ref{fig3} (a), $\alpha_1(z)$ and  $\alpha_2(z)$  are defined  to be the angles between $\hat{\mathbf{e}}_{z}$ and the incident directions from the lower and upper ends of $\mathcal L_s$ to the point $z$.  It is trivial to see that for any choices of $\mathbf{\Omega}$, 
\begin{align}\label{eq:spatial_bandwidth_wz}
	w_{z}(z; \mathbf{\Omega}) 
	&= \lambda^{-1} [ \cos(\alpha_1(z) ) -  \cos(\alpha_2(z))] \nonumber \\
	& =   \frac{1}{\lambda} \frac{(z+a)}{\sqrt{(z+a)^2 + d^2}} -  \frac{1}{\lambda} \frac{z+b}{\sqrt{(z+b)^2 + d^2}} ,	 
\end{align}
where 
\begin{equation}
		d= r \sin\theta , \quad
		a = r \cos\theta + \frac{L}{2} , \quad
		b = r \cos\theta - \frac{L}{2}.
\end{equation}
By examining \eqref{eq:spatial_bandwidth_wz} or the geometric relation shown by Fig. \ref{fig3}~(a), it can be seen that 
$$w_{z}(z; L, r,\theta) = w_{z}(- z; L, r,\pi-\theta).	$$
	
We aim to derive the explicit expressions of  $w_{z}^{\mathrm{max}} (\overline{\mathbf{\Omega}})$ and $w_{z}^{\mathrm{min}} ( \overline{\mathbf{\Omega}})$. The above symmetric relation allows us to first focus on the case of $\theta \in [0, \pi/2]$ and thus $\cos\theta \geq 0$, and then extend the result to the entire range of $\theta$. If we relax the valid range  of $z$ to be $(-\infty , \infty)$, it can be verified that $w_{z}(z; \mathbf{\Omega})$ is maximized at 
	\begin{equation}
		z_0 = -\frac{a+b}{2} = -r\cos\theta, 
	\end{equation}
where the derivative of $w_{z}(z; \mathbf{\Omega})$, given by 
	\begin{equation}
		\frac{\mathrm{d} w_{z}(z; \mathbf{\Omega}) }{\mathrm{d} z} 
		= \frac{d^2}{\lambda} \! \left( (z+a)^2  + d^2 \right)^{-\frac{3}{2}} - \frac{d^2}{\lambda}\! \left( (z+b)^2 + d^2 \right)^{-\frac{3}{2}}, 
	\end{equation}
equals to $0$. Moreover, 
	\begin{equation}\label{eq:sw_z_max0}
		w_{z}^{\mathrm{max},0}(\mathbf{\Omega})   \triangleq 	w_{z}(z_0; \mathbf{\Omega}) =  \frac{1}{\lambda}  \frac{ L }{\sqrt{ L^2/4  + d^2}} . 
	\end{equation}
is the maximum possible spatial bandwidth that can be seen on $\mathcal L_z$. It is not difficult to see that  $w_{z}(z)$  is symmetric with respect to $z = z_0$. However, $z_0$ falls within the valid range of $z$ only when $r\cos\theta  \leq \rho$. Otherwise, $w_{z}(z)$ is a monotonically decreasing function over $z\in [-\rho, \rho]$ as $z_0 < -\rho$.
	
Based on the symmetry in $w_{z}(z)$ and the discussions above, we have
\begin{align}
	w_{z}^{\mathrm{max}}(\overline{\mathbf{\Omega}}) &= \begin{cases}
			w_{z}( \rho; \mathbf{\Omega}) ,  &  r \cos\theta < - \rho, \\
			w_{z}^{\mathrm{max},0}(\mathbf{\Omega}) ,   & -\rho \leq   r \cos\theta \leq  \rho, \\
			w_{z}( -\rho; \mathbf{\Omega}) ,  &  r \cos\theta> \rho,
		\end{cases} \label{eq:sw_z_max1} \\
	w_{z}^{\mathrm{min}}(\overline{\mathbf{\Omega}}) &= \begin{cases}
			w_{z}( -\rho; \mathbf{\Omega}) ,  &  r \cos\theta \leq 0, \\
			w_{z}( \rho; \mathbf{\Omega}) ,  &  r \cos\theta > 0. 
		\end{cases} \label{eq:sw_z_min1}
\end{align}
These results can also be obtained through geometric analysis. In particular, we would like to point out that when $-\rho \leq r \cos\theta \leq  \rho$,  the orthogonal projection of  the center of $\mathcal L_s$  onto $\mathcal L_z$ is exactly at $z = z_0$. It is then easy to prove that $w_{z}^{\mathrm{max},0} = \cos(\alpha_1^0 ) -  \cos(\alpha_2^0)$ is the maximum possible spatial bandwidth, where $\alpha_1^0 $ and $\alpha_2^0$ are the angles between $\hat{\mathbf{e}}_{z}$ and the incident directions from the lower and upper ends of $\mathcal L_s$ to this particular point. 

To make the expressions more compact, we define 
	\begin{equation}\label{eq:sw_f}
		f(t; c) =  \frac{1}{\lambda} \frac{t }{\sqrt{t^2 + c^2}}, 
	\end{equation} 
and
\begin{equation}\label{eq:sw_AB}
	A = r|\cos\theta|+\frac{L}{2} , \quad  
	B = r|\cos\theta|-\frac{L}{2} .
\end{equation}
Equations \eqref{eq:sw_z_max1} and \eqref{eq:sw_z_min1} can then be rewritten as follows:  
\begin{align}
	w_{z}^{\mathrm{max}}(\overline{\mathbf{\Omega}}) &= \begin{cases}
			2\, f(\frac{L}{2} ; d)  ,   &  |\cos\theta| \leq  \frac{\rho}{r}, \\
			f(A -\rho; d) - f(B -\rho; d) ,  &   |\cos\theta| > \frac{\rho}{r}, 
		\end{cases}\label{eq:sw_z_max} \\
		w_{z}^{\mathrm{min}}(\overline{\mathbf{\Omega}}) &= f(A +\rho; d) - f(B+\rho; d ). \label{eq:sw_z_min}
\end{align}

Following \eqref{eq:spatial_DoF_RCS_apx2}, the linear approximation formula for the K number achieved by $\mathcal L_z$ is 
	\begin{equation}\label{eq:DoF_lin_apx_z} 
		{K}^{a}_{z}(\overline{\mathbf{\Omega}} )  
		= \rho \left( w_{z}^{\mathrm{max}}(\overline{\mathbf{\Omega}}) +    w_{z}^{\mathrm{min}}(\overline{\mathbf{\Omega}}) \right) .
	\end{equation}
The explicit expression of $w_{z}^{\mathrm{range}} (\overline{\mathbf{\Omega}}) = w_{z}^{\mathrm{max}}(\overline{\mathbf{\Omega}}) -    w_{z}^{\mathrm{min}}(\overline{\mathbf{\Omega}})$ is also trivial to obtain.
	
\subsection{In \texorpdfstring{$ \hat{\mathbf{e}}_{x} $}{ex} Direction} 
\label{sec:5b}
	
Denote $w_{x}(x;  \mathbf{\Omega})$ to be the local spatial bandwidth of the electric field perceived at a point on $\mathcal L_x$ specified by coordinate $x$, as depicted in Fig. \ref{fig3} (b). First of all, we identify the effective integration range for $w_{x}(x;  \mathbf{\Omega})$, denoted by $\mathcal I_x(\overline{\mathbf{\Omega}} )$. When $d< \rho$, care should be taken as a smaller part of $\mathcal L_x$ (with $-\rho \leq x < -d$) will lie on the different half of the $\mathrm{o}$-$\mathcal L_s$ plane than the larger part. Owing to the geometric symmetry, the electric field on this smaller part is a mirror image of the other side and contributes no extra spatial DoF.  As a result, the effective  integration range can be expressed as $\mathcal I_x(\overline{\mathbf{\Omega}} ) =  [-\min\{ d, \rho \} ,  \rho]$ to cover all circumstances. We note that $d < \rho$ happens only when $\mathcal L_x$ is very close to the $\mathrm z'$-axis and $\rho$ is sufficiently large. Such geometric conditions may  have limited practical relevance.
	
As shown in  Fig. \ref{fig3} (b), $\beta_1(x)$ and  $\beta_2(x)$ are defined as the angles between $\hat{\mathbf{e}}_{x}$ and the incident directions from the lower and upper ends of $\mathcal L_s$ to the point $x$, and we have 
	\begin{align}
		\cos(\beta_1(x) ) =  \frac{x+d}{\sqrt{(x+d )^2 + a^2}},  \\
		\cos(\beta_2(x) ) =  \frac{x+d}{\sqrt{(x+d )^2 + b^2}}. 
	\end{align}
When the orthogonal projection of the center of $\mathcal L_x$ onto the $\mathrm z'$-axis falls on $\mathcal L_s$, the incident direction from this  point contributes to the smallest angle, which is $0$ since it coincides with the $\mathrm{x}$-axis. Meanwhile, the end of $\mathcal L_s$ that is closer to $\mathrm{o}$ contributes to the largest angle. Otherwise,  $\beta_1(x)$ and  $\beta_2(x)$ are the two limits. Accordingly,  it can be concluded that 
\begin{align}
\label{eq:spatial_bandwidth_wx}
	&w_{x}(x; \mathbf{\Omega}) = \nonumber \\
		& \; \begin{cases}
			\frac{1}{\lambda} [\cos(\beta_1(x))  -   \cos(\beta_2(x))]   ,  &  r \cos\theta < -\frac{L}{2},\\
			\frac{1}{\lambda} [ 1 -  \cos(\beta_2(x))] ,    & - \frac{L}{2}  \leq  r \cos\theta  \leq 0,\\
			\frac{1}{\lambda} [1 -  \cos(\beta_1(x)) ] ,   & 0 <  r \cos\theta  \leq  \frac{L}{2}, \\
			\frac{1}{\lambda} [\cos(\beta_2(x) ) -  \cos(\beta_1(x))] ,  & r \cos\theta> \frac{L}{2}. 
		\end{cases}
\end{align}	
We remark  that $\cos(\beta_1(x) ) \geq 0$, $\cos(\beta_2(x) ) \geq 0$, and $ w_{x}(x)  > 0$ always hold within the valid range of $x$, which ensures  $x+d\geq 0$. Moreover, the following symmetric relation can be easily seen:
$$w_{x}(x; L, r,\theta) = w_{x}(x; L, r,\pi-\theta).$$
Adopting \eqref{eq:sw_f} and \eqref{eq:sw_AB}, $w_{x}(x; \mathbf{\Omega})$ can also be given in a more compact form:
\begin{align}
\label{eq:spatial_bandwidth_wx1}
	w_{x}(x; \mathbf{\Omega}) = 
		\begin{cases}
			\frac{1}{\lambda}  -  f(x+d; A) ,   &  |\cos\theta| \leq  \frac{L}{2r}, \\
			f(x+d; B) - f(x+d; A),  & |\cos\theta| > \frac{L}{2r}. 
		\end{cases}
\end{align}

To derive explicit expressions of  $w_{x}^{\mathrm{max}} (\overline{\mathbf{\Omega}})$ and $w_{x}^{\mathrm{min}} ( \overline{\mathbf{\Omega}})$, we again focus on the case of  $\theta \in [0, \pi/2]$ first and then extend the result to the entire range using symmetry. When  $0\leq \cos\theta  \leq \frac{L}{2r} $,  $w_{x}(x) $ is a decreasing function of $x$ within $\mathcal I_x(\overline{\mathbf{\Omega}} )$, and therefore, $w_{x}^{\mathrm{max}}  = w_{x}(-\min\{ d, \rho \}  )$.  When $ \cos\theta > \frac{L}{2r}$, the maximum spatial bandwidth is seen at the point $x = x_0$, where the derivative of $w_{x}(x)$ equals to zero, provided that $x_0$ falls in the valid range of $x$.  In particular, 
	\begin{align}
		\frac{\mathrm{d} w_{x}(x; \mathbf{\Omega})}{\mathrm{d} x}  
		= \frac{a^2}{\lambda}\! \left( a^2 \!+ \!(x+d)^2 \right)^{-\frac{3}{2}} \!- \frac{b^2}{\lambda}\! \left( b^2 +\! (x+d)^2 \right)^{-\frac{3}{2}}, 
	\end{align}
which leads to 
	\begin{equation}\label{eq:sw_x_x0}
		x_0 =  \sqrt{ \frac{ a^{\frac{2}{3}} - b^{\frac{2}{3}} }{ b^{-\frac{4}{3}} - a^{- \frac{4}{3}} }  } - d.
	\end{equation}
It should be noted that $\cos(\beta_1(x))  -  \cos(\beta_2(x))$ is not a symmetric function with respect to $x = x_0$. As a result, it is nontrivial to determine which end of $\mathcal L_x$ sees a smaller spatial bandwidth when $x_0 \in \mathcal I_x(\overline{\mathbf{\Omega}} )$. When  $x_0$ falls out of $\mathcal I_x(\overline{\mathbf{\Omega}} )$, $ w_{x}(x)$ becomes monotonic within $\mathcal I_x(\overline{\mathbf{\Omega}} )$. Moreover, it is increasing if $x_0 > \rho$ and decreasing if $ x_0 < -\min\{ d, \rho \} $.  
	
Based on the above discussions, using \eqref{eq:sw_f} and \eqref{eq:sw_AB}, we summarize the explicit expressions for  $w_{x}^{\mathrm{max}}(\overline{\mathbf{\Omega}})$ and  $w_{x}^{\mathrm{min}}(\overline{\mathbf{\Omega}})$  in \eqref{eq:sw_x_max} and \eqref{eq:sw_x_min} respectively, {shown at the bottom of the next page}. The operator $(\cdot)^+ \triangleq \max(\cdot, 0)$ is adopted.

\begin{figure*}[b]
\hrule
\begin{equation}
\label{eq:sw_x_max}
w_{x}^{\mathrm{max}}(\overline{\mathbf{\Omega}}) = \begin{cases}
	\frac{1}{\lambda} -  f( (d-\rho)^+; A)  ,   &  |\cos\theta| \leq  \frac{L}{2 r}, \\
	f( (d-\rho)^+; B) -   f( (d-\rho)^+; A)  ,  & |\cos\theta| > \frac{L}{2r}, \, x_0 < -\min\{ d, \rho \}  , \\
	f(d+x_0; B) -   f(d+x_0; A)  ,  &|\cos\theta| > \frac{L}{2r}, \, -\min\{ d, \rho \} \leq x_0 \leq \rho, \\
	f(d+\rho; B) - f(d+\rho; A) ,  & |\cos\theta| > \frac{L}{2r},\, x_0 > \rho.  
	\end{cases}
\end{equation}	
\begin{equation}
	\label{eq:sw_x_min}
w_{x}^{\mathrm{min}}(\overline{\mathbf{\Omega}}) = \begin{cases}
	\frac{1}{\lambda} -  f(d+\rho; A)  ,   &  |\cos\theta| \leq  \frac{L}{2 r}, \\
	f( d+\rho; B) -   f( d+\rho; A)  ,  & |\cos\theta| > \frac{L}{2r},  x_0 < -\min\{ d, \rho \}  , \\
	\min \big\{ f( (d-\rho)^+; B) -  f( (d-\rho)^+; A) , f(d+\rho; B) - f(d+\rho; A) \big\}  ,  &|\cos\theta| > \frac{L}{2r},  -\min\{ d, \rho \} \leq x_0 \leq \rho, \\
	f((d-\rho)^+; B) - f((d-\rho)^+; A) ,  & |\cos\theta| > \frac{L}{2r}, x_0 > \rho.  
	\end{cases}
\end{equation} 
\end{figure*}	
	
Following \eqref{eq:spatial_DoF_RCS_apx2}, the linear approximation formula for the K number  achieved by $\mathcal L_x$ is as follows:
\begin{equation}\label{eq:DoF_lin_apx_x} 
	{K}^{a}_{x}(\overline{\mathbf{\Omega}} )  = 
	\frac{(\rho + \min\{ d, \rho \} ) }{2}  
	\left( w_{x}^{\mathrm{max}}(\overline{\mathbf{\Omega}}) +   w_{x}^{\mathrm{min}}(\overline{\mathbf{\Omega}}) \right)  . 
\end{equation}
Using \eqref{eq:sw_x_max} and \eqref{eq:sw_x_min}, the explicit expression of $w_{x}^{\mathrm{range}} (\overline{\mathbf{\Omega}}) =  w_{x}^{\mathrm{max}}(\overline{\mathbf{\Omega}}) -  w_{x}^{\mathrm{min}}(\overline{\mathbf{\Omega}})$ is trivial to obtain.

\subsection{In \texorpdfstring{$ \hat{\mathbf{e}}_{y} $}{ey} Direction} 
\label{sec:5c}
	
Denote $w_{y}(y; \mathbf{\Omega})$ to be the local spatial bandwidth of the  electric field seen at a point on $\mathcal L_y$ specified by coordinate $y$, as depicted in Fig. \ref{fig3} (c). Due to the way we define the local coordinate system, the electric field seen along $\mathcal L_y$ is perfectly symmetric with respective to $y=0$. Therefore, the effective integration range for $w_{y}(y;  \mathbf{\Omega})$ can be chosen either as $\mathcal I_y^+(\overline{\mathbf{\Omega}} ) = [0, \rho]$ or as $\mathcal I_y^-(\overline{\mathbf{\Omega}} ) = [-\rho, 0]$ for any given $\mathbf{\Omega}$. 
	
To cover both choices, we define $\eta_1(y)$ and  $\eta_2(y)$ to be the angles between the incident directions from the lower and upper ends of $\mathcal L_s$ to the point $y $ and  $\hat{\mathbf{e}}_{y}$ when choosing $\mathcal I_y^+(\overline{\mathbf{\Omega}} )$, or $- \hat{\mathbf{e}}_{y}$ when choosing $ \mathcal I_y^-(\overline{\mathbf{\Omega}} )$. Fig. \ref{fig3} (c) illustrates the angles under the choice of $\mathcal I_y^+(\overline{\mathbf{\Omega}} )$.  As a result, we have 
\begin{align}
	\cos \eta_1(y)  =  \frac{|y|}{\sqrt{d^2 + a^2 + y^2 }} , \\
	\cos \eta_2(y)  =   \frac{|y|}{\sqrt{d^2 + b^2 + y^2 }}. 
\end{align}
Similar to the $\hat{\mathbf{e}}_{x}$ case, when the orthogonal projection of $\mathrm{o}$ onto the  $\mathrm z'$-axis falls on $\mathcal L_s$, the incident direction from this projection point contributes to the smallest angle $\eta_0(y)$, defined in a way similar to $\eta_1(y)$ and $\eta_2(y)$, and 
\begin{equation}
	\cos\eta_0(y) = \frac{|y|}{\sqrt{d^2 + y^2 }} .
\end{equation}
In this case, the end of  $\mathcal L_s$ that is closer to $\mathrm{o}$ contributes to the largest angle. Otherwise, the angle is bounded by $\eta_1(y)$ and  $\eta_2(y)$. Accordingly, we have for $y\in [-\rho, \rho]$, 
\begin{equation}\label{eq:spatial_bandwidth_wy}
	w_{y}(y;\mathbf{\Omega})  = \begin{cases}
			\frac{1}{\lambda} [\cos\eta_1(y)  - \cos\eta_2(y)] ,  &  r \cos\theta < -\frac{L}{2},\\
			\frac{1}{\lambda} [\cos\eta_0(y)  - \cos\eta_2(y) ] ,    & - \frac{L}{2}  \leq  r \cos\theta  \leq 0,\\
			\frac{1}{\lambda} [\cos\eta_0(y)  - \cos\eta_1(y)] ,   & 0 <  r \cos\theta  \leq  \frac{L}{2}, \\
			\frac{1}{\lambda} [\cos\eta_2(y)  - \cos\eta_1(y)] ,  & r \cos\theta> \frac{L}{2}.
		\end{cases}
\end{equation}	

For $w_{y}(y;\mathbf{\Omega})$, we have the following symmetric relations
\begin{align*}
   & w_{y}(-y; L, r,\theta) = w_{y}(y; L, r,\theta), \\
   & w_{y}(y; L, r,\theta) = w_{y}(y; L, r,\pi-\theta). 
\end{align*}
Moreover, by adopting \eqref{eq:sw_f} and \eqref{eq:sw_AB}, the expression of $w_{y}(y;\mathbf{\Omega})$ can also be simplified as 
\begin{align}
    & w_{y}(y;\mathbf{\Omega})  = \nonumber \\ 
    &\; \begin{cases}
		f(|y|;d) - f(|y|; \sqrt{d^2 + A^2})  ,   & |\cos\theta | \leq  \frac{L}{2r}, \\
			f(|y|; \sqrt{d^2 + B^2}) - f(|y|; \sqrt{d^2 + A^2}),  &  |\cos\theta | > \frac{L}{2r}.
	\end{cases}
\end{align}

Since $\hat{\mathbf{e}}_{y}$ is perpendicular to the $\mathrm{o}$-$\mathcal L_s$ plane, the minimum value of $w_{y}(y; \mathbf{\Omega})$ appears at $y=0$ regardless of the choices of $\mathbf{\Omega}$, and in particular, 
\begin{equation}
	w_{y}^{\mathrm{min}} (\overline{\mathbf{\Omega}})    = w_{y}(0; \mathbf{\Omega}) =0 .
\end{equation}
Moreover, the maximum spatial bandwidth always appears at $y= \pm \rho$. As a result, 
	\begin{equation}
		w_{y}^{\mathrm{max}} (\overline{\mathbf{\Omega}})  \equiv w_{y}^{\mathrm{range}} (\overline{\mathbf{\Omega}}) = w_{y}(\rho; \mathbf{\Omega}) . 
	\end{equation}
Following \eqref{eq:spatial_DoF_RCS_apx2}, the linear approximation formula for the K number  achieved by $\mathcal L_z$ is as follows 
	\begin{equation}\label{eq:DoF_lin_apx_y} 
		{K}^{a}_{y}(\overline{\mathbf{\Omega}} )  = \frac{\rho }{2}  w_{y}^{\mathrm{max}}(\overline{\mathbf{\Omega}}). 
	\end{equation}
Finally, we remark that the spatial DoF available in this direction actually comes from the change in the radial distance between the observing positions and $\mathcal L_s$.

\section{Spatial-Multiplexing Region Boundaries}
\label{sec:6}
	
Based on the analytical results obtained in the previous section, we examine the boundaries of the spatial multiplexing regions of the three elementary linear array assemblies.  
	
\subsection{In \texorpdfstring{$\hat{\mathbf{e}}_{z}$}{ez}  and \texorpdfstring{$\hat{\mathbf{e}}_{x}$}{ez} Directions} 	
\label{sec:6a}
	
\subsubsection{Spatial multiplexing region boundaries}
	
To find the spatial multiplexing region boundaries for these two receiving directions, we need to solve the two equations $K_{z}(\overline{\mathbf{\Omega}} ) = K_0$ and $K_{x}(\overline{\mathbf{\Omega}} ) = K_0$, treating $r$ as the only unknown in $\overline{\mathbf{\Omega}}$, for all $\theta \in [0,\pi]$. Due to the involved integration, these two equations are not straightforward to solve. It can be inferred from \eqref{eq:spatial_bandwidth_wz} and \eqref{eq:spatial_bandwidth_wx} that when $r$ is sufficiently large, which is the case if a small $K_0$ is targeted, the change in the local spatial bandwidth over $\mathcal L_z$ and $\mathcal L_x$ is little, and  $w_{z}^{\mathrm{max}}(\overline{\mathbf{\Omega}})$ and $w_{x}^{\mathrm{max}}(\overline{\mathbf{\Omega}})$ are good constant approximations. Therefore, we replace $K_{z}(\overline{\mathbf{\Omega}} )$ using  $K^{\mathrm u}_{z}(\overline{\mathbf{\Omega}} )$, and $K_{x}(\overline{\mathbf{\Omega}} )$ using $K^{\mathrm u}_{x}(\overline{\mathbf{\Omega}} )$, to make the above two equations more tractable. 
	
Following the discussion in Section \ref{sec:5a}, $K^{\mathrm u}_{z}(\overline{\mathbf{\Omega}} ) = 2\rho\,w_{z}^{\mathrm{max}} (\overline{\mathbf{\Omega}})$ is immediately obtained. Since $\mathcal I_x(\overline{\mathbf{\Omega}} ) \neq [-\rho, \rho]$ only under very extreme geometric conditions, as discussed in Section \ref{sec:5b}, we adopt $[-\rho, \rho]$ to be the effective integration range for $\mathcal L_x$ under all geometric conditions as well for simplicity and without significant loss of accuracy. Namely, the following approximation is adopted: 
	\begin{equation}\label{eq:DoF_max_apx_x}
	    K^{\mathrm u}_{x}(\overline{\mathbf{\Omega}} ) \approx 2\rho \,w_{x}^{\mathrm{max}} (\overline{\mathbf{\Omega}}) . 
	\end{equation}
  
Accordingly, we determine the boundaries of spatial multiplexing regions $\mathcal R_{z}(\overline{\mathbf{\Omega}}, K_0)$ and $\mathcal R_{x}(\overline{\mathbf{\Omega}}, K_0)$ by solving the following two equations instead: 
	\begin{align}\label{eq:PW_zx}
		w_{z}^{\mathrm{max}}(\overline{\mathbf{\Omega}})  = \frac{{K_0}}{2\rho}, \quad 
		w_{x}^{\mathrm{max}}(\overline{\mathbf{\Omega}})  = \frac{{K_0}}{2\rho}. 
	\end{align}
The resulting distance threshold at polar angle $\theta$ will be denoted by $R^{\mathrm u}_{z}(\theta ;  L, \rho, K_0)$ and $R^{\mathrm u}_{x}(\theta ;  L, \rho, K_0)$ in the following discussions. 
	
\subsubsection{Distance threshold \texorpdfstring{$R_0$}{R0}}
	
The most favorable geometric relation for the availability of spatial DoF is when $\mathcal L_z$ is located in the boresight direction of $\mathcal L_s$, namely, when $\theta = \pi/2$. Following the discussion in Section \ref{sec:5a}, the distance threshold of the spatial multiplexing region in this direction, namely, $R_z(\pi/2; L,\rho,K_0)$,  is given by the solution of the following equation:
	\begin{equation}
		w_{z}^{\mathrm{max},0}(r; L, \rho, \theta = \pi/2)  =   \frac{1}{\lambda} \frac{ L }{\sqrt{ L^2/4 + r^2}}  =  \frac{K_0}{2\rho}. 
	\end{equation}
Note that $r \cos\theta = 0$ and $d= r \sin \theta = r $ are substituted into \eqref{eq:sw_z_max0} to formulate the above equation. It is easily derived that 
	\begin{equation}\label{eq:Rz_K0} 
		R_z(\pi/2; L,\rho,K_0) =  \frac{L}{\lambda}  \sqrt{ \frac{4\rho^2}{K_0^2} - \frac{1}{4}}. 
	\end{equation}
	
In addition, we define $R_0(L,\rho)$ as the distance threshold of $K_0 =1$ under the same conditions, namely, 
	\begin{equation}\label{eq:R0_K1} 
		R_0(L,\rho) \triangleq R_z(\pi/2; L,\rho,K_0 =1) = \frac{L}{\lambda}  \sqrt{ 4\rho^2 - \frac{1}{4} }.
	\end{equation}
It is the maximum distance threshold of $K_0=1$ of the spatial multiplexing region of a linear receiving array in all direction and orientation conditions. If $\rho$ is not too small, the following approximation holds:  
	\begin{equation}\label{eq:R0_max} 
		R_0 (L,\rho)  \approx \frac{2\rho L }{\lambda}  .
	\end{equation}
We note that this is exactly the same result obtained from  \eqref{eq:Kparallel_1} (for $D$) by requiring  $K_{\mathrm{parallel}} = 1$.

\begin{remark}
There is a tendency in some recent literature to use the Fraunhofer distance, which is given by $D_{\mathrm{F}}(L) = \frac{L^2}{\lambda}$ and usually used for dividing the near and far radiation fields of an antenna array \cite{selvan2017fraunhofer}, as the availability criterion for spatial multiplexing capability. 
In the spatial region bounded by $D_{\mathrm{F}}$, often it is necessary to consider the actual spherical wavefront to avoid phase computation errors and performance loss in array-based applications  \cite{jiang2005spherical, zhou2015spherical}. We note that if $2\rho = L$, then $D_{\mathrm{F}}(L) = R_0 (L,\rho) $.  Therefore, the Fraunhofer distance becomes the distance threshold for the spatial multiplexing region, provided that the source and receiving arrays are parallel, of similar size, and located in the boresight direction of each other. In the general case, however, the Fraunhofer distance alone does not imply spatial multiplexing capability in the LOS channel. 
\end{remark} 
	
\subsubsection{Non-constant-bandwidth spatial multiplexing region boundaries}
	
Following \eqref{eq:CBSMR}, the boundaries of $\mathcal R^{\mathrm{nc}}_{z}(\overline{\mathbf{\Omega}}, \Delta K)$ and $\mathcal R^{\mathrm{nc}}_{x}(\overline{\mathbf{\Omega}}, \Delta K)$ can be found by solving $K^u_{z}(\overline{\mathbf{\Omega}} ) - K_{z}(\overline{\mathbf{\Omega}} ) =  \Delta K$ and $K^u_{x}(\overline{\mathbf{\Omega}} ) - K_{x}(\overline{\mathbf{\Omega}} ) =  \Delta K$, which is again difficult due to the involved integration.  It will be shown later that  $K^{\mathrm a}_{z}(\overline{\mathbf{\Omega}} )$ and $K^{\mathrm a}_{x}(\overline{\mathbf{\Omega}} )$, given by \eqref{eq:DoF_lin_apx_z} and \eqref{eq:DoF_lin_apx_x}, closely approximate  $K_{z}(\overline{\mathbf{\Omega}} ) $ and $K_{x}(\overline{\mathbf{\Omega}})$, respectively. Therefore, they are adopted to replace $K_{z}(\overline{\mathbf{\Omega}} )$ and $K_{x}(\overline{\mathbf{\Omega}} )$ in the above equations for better tractability.  Moreover, following the same consideration as discussed before \eqref{eq:PW_zx},  we again adopt $[-\rho, \rho]$ as the effective integration range for  $w_{x}(x;  \mathbf{\Omega})$ under all geometric conditions. Namely,  ${K}^{a}_{x}(\overline{\mathbf{\Omega}} )$ given by \eqref{eq:DoF_lin_apx_x} is further approximated as 
	\begin{equation}\label{eq:DoF_lin_apx_x2}
		{K}^{a}_{x}(\overline{\mathbf{\Omega}} )  \approx \rho  \left( w_{x}^{\mathrm{max}}(\overline{\mathbf{\Omega}}) + w_{x}^{\mathrm{min}}(\overline{\mathbf{\Omega}}) \right). 
	\end{equation} 
	
Substituting $K^{\mathrm u}_{z}(\overline{\mathbf{\Omega}} ) = 2\rho\,w_{z}^{\mathrm{max}} (\overline{\mathbf{\Omega}})$, and $K^u_{x}(\overline{\mathbf{\Omega}} )$, $K^a_{z}(\overline{\mathbf{\Omega}} )$, and  $K^a_{x}(\overline{\mathbf{\Omega}} )$ using \eqref{eq:DoF_max_apx_x}, \eqref{eq:DoF_lin_apx_z}, and \eqref{eq:DoF_lin_apx_x2}, respectively, and recall the definition of $w_{z}^{\mathrm{range}}(\overline{\mathbf{\Omega}})$ and $w_{x}^{\mathrm{range}}(\overline{\mathbf{\Omega}})$, the equations we need to solve are immediately obtained: 
\begin{align}\label{eq:SW_zx}
	w_{z}^{\mathrm{range}}(\overline{\mathbf{\Omega}})  =  \frac{{\Delta K}}{\rho}, \quad 
		w_{x}^{\mathrm{range}}(\overline{\mathbf{\Omega}})  =  \frac{{\Delta K}}{\rho}. 
\end{align}
Again, $r$ is treated as the only unknown in $\overline{\mathbf{\Omega}}$ and the equations are solved for all $\theta \in [0,\pi]$. We note that that multiple solutions may exist for certain $\theta$, which will also be shown later using numerical results. The resulting distance thresholds at polar angle $\theta$ are denoted by $R_z^{\mathrm{nc,a}}(\theta; L,\rho,\Delta K)$ and $R_x^{\mathrm{nc,a}}(\theta; L,\rho,\Delta K)$ respectively. 
	
\subsection{In \texorpdfstring{$\hat{\mathbf{e}}_{y}$}{ey} Direction} 	
\label{sec:6b}
	
As discussed in Section \ref{sec:5c}, when the linear receiving array is orientated in the $\hat{\mathbf{e}}_{y}$ direction, the minimum value (which is $0$) of the local spatial bandwidth always appears at the center and the maximum value is seen at either end of the array. Therefore, the local spatial bandwidth distributed over $\mathcal L_y$ should not be approximated as a constant under this orientation condition, unless the maximum value is approximately zero.  
Accordingly, we stress that \textit{the entire spatial multiplexing region for the $\hat{\mathbf{e}}_{y}$ receiving direction should be regarded as the non-constant bandwidth spatial multiplexing region.}  Therefore, the linear approximation  ${K}^{a}_{y}(\overline{\mathbf{\Omega}} )$ given by \eqref{eq:DoF_lin_apx_y} is adopted to replace the actual K number ${K}_{y}(\overline{\mathbf{\Omega}} )$, and the equation ${K}^{a}_{y}(\overline{\mathbf{\Omega}} ) = K_0$ is to be solved to find the boundary of this region. Substituting \eqref{eq:DoF_lin_apx_y}, this equation translates immediately to 
	\begin{align}\label{eq:SW_y}
		w_{y}^{\mathrm{max}}(\overline{\mathbf{\Omega}})  =  \frac{{2 K_0}}{\rho}.
	\end{align} 
The resulting distance threshold at polar angle  $\theta$ is denoted by denoted by $R^{\mathrm a}_{y}(\theta ;  L, \rho, K_0)$ in the following discussion.

\begin{remark} 
	Given the generic receiving array $\mathcal L_r$ orientated in $\hatbf v$ direction, the boundaries obtained following the same steps as described above, but using the actual lengths of its projections in the three orthogonal receiving directions instead of $2\rho$, shall be used to examine if these directions are contributing to the achievable spatial DoF or not. 
\end{remark}

\subsection{Numerical Results in A Case Study}
\label{sec:6c}
	
In this subsection, we present some numerical results obtained under the setting of $L = 400 \lambda$ and $\rho = 20\lambda$, which could be considered as a LOS communication scenario between an LSAA and a smaller receiving array (of length $40 \lambda$). $K_0 = 1$ and $\Delta K  = 1$ are adopted to determine the boundaries of the spatial multiplexing regions. Following \eqref{eq:R0_max}, the distance threshold $R_0(L,\rho)$ is given by $1.6\times 10^4 \lambda$.

\begin{figure}[!t]
		\centering
		\subfigure[$R^{\mathrm u}_{z}(\theta ;  L, \rho,K_0=1)$]{\includegraphics [width= .47\linewidth,trim= 95 115 110 110, clip]{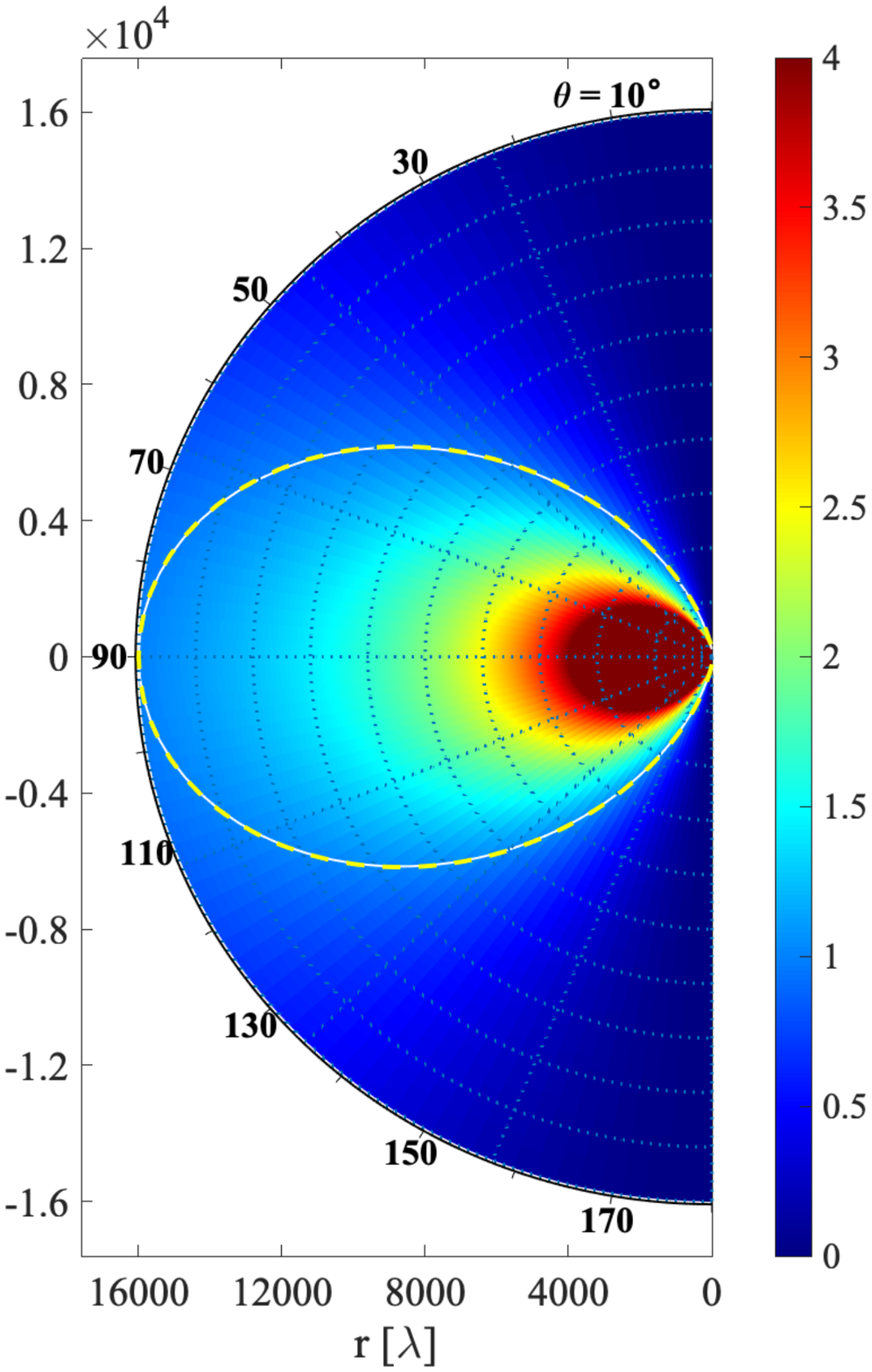} }
		\subfigure[$R^{\mathrm u}_{x}(\theta ;  L, \rho,K_0=1)$]{\includegraphics [width= .47\linewidth,trim= 95 115 110 110, clip]{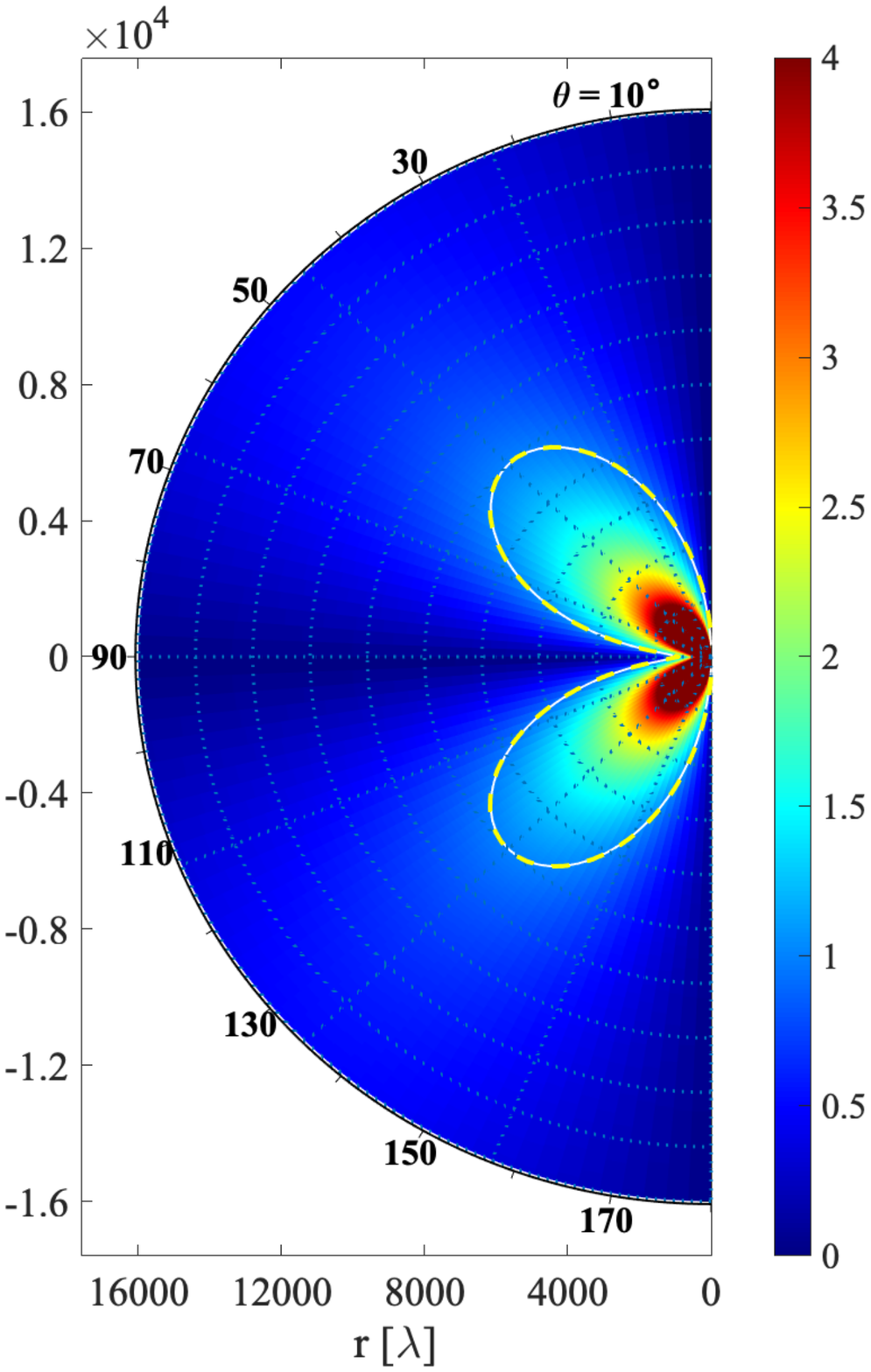} }
		\caption{Spatial multiplexing region boundaries $R^{\mathrm u}_{z}(\theta ;  L, \rho,K_0=1)$ and $R^{\mathrm u}_{x}(\theta ;  L, \rho,K_0=1)$ (in dashed yellow lines) presented over the K number color maps, for $r \in [0, R_0(L,\rho) ]$ and $\theta \in [0, \pi]$, with $L = 400\lambda$ and $\rho = 20\lambda$.  The white contour lines (mostly overlapped with the dashed yellow lines) represent the corresponding exact boundaries.}
		\label{fig4}
	\end{figure}

\begin{figure}[!t]
		\centering
		\subfigure[$R^{\mathrm nc,a}_{z}(\theta ;  L, \rho,\Delta K=1)$]{\includegraphics [width= .47\linewidth,trim= 95 115 110 110, clip]{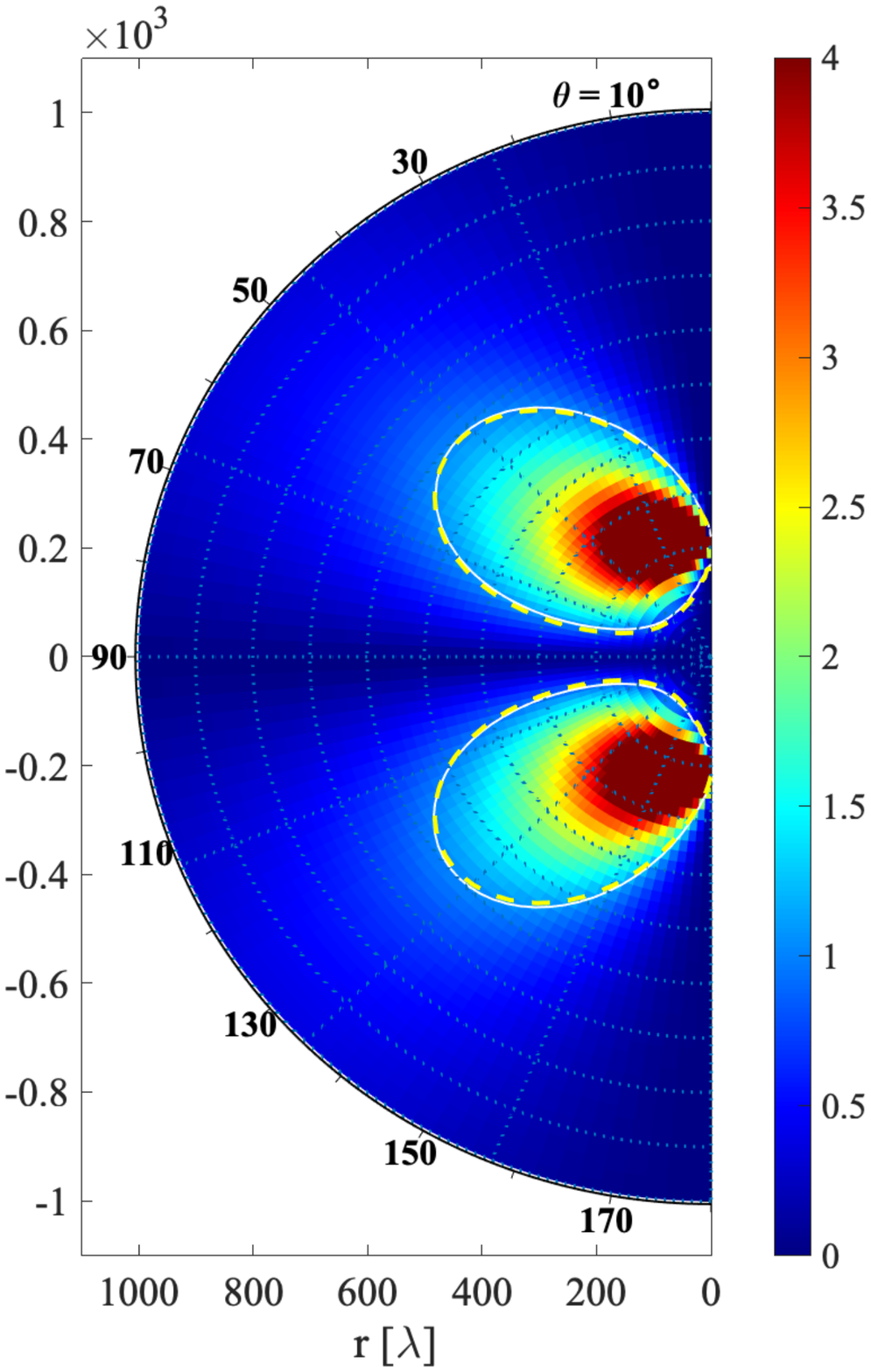} }
		\subfigure[$R^{\mathrm nc,a}_{x}(\theta ;  L, \rho,\Delta K=1)$]{\includegraphics [width= .47\linewidth,trim= 95 115 110 110, clip]{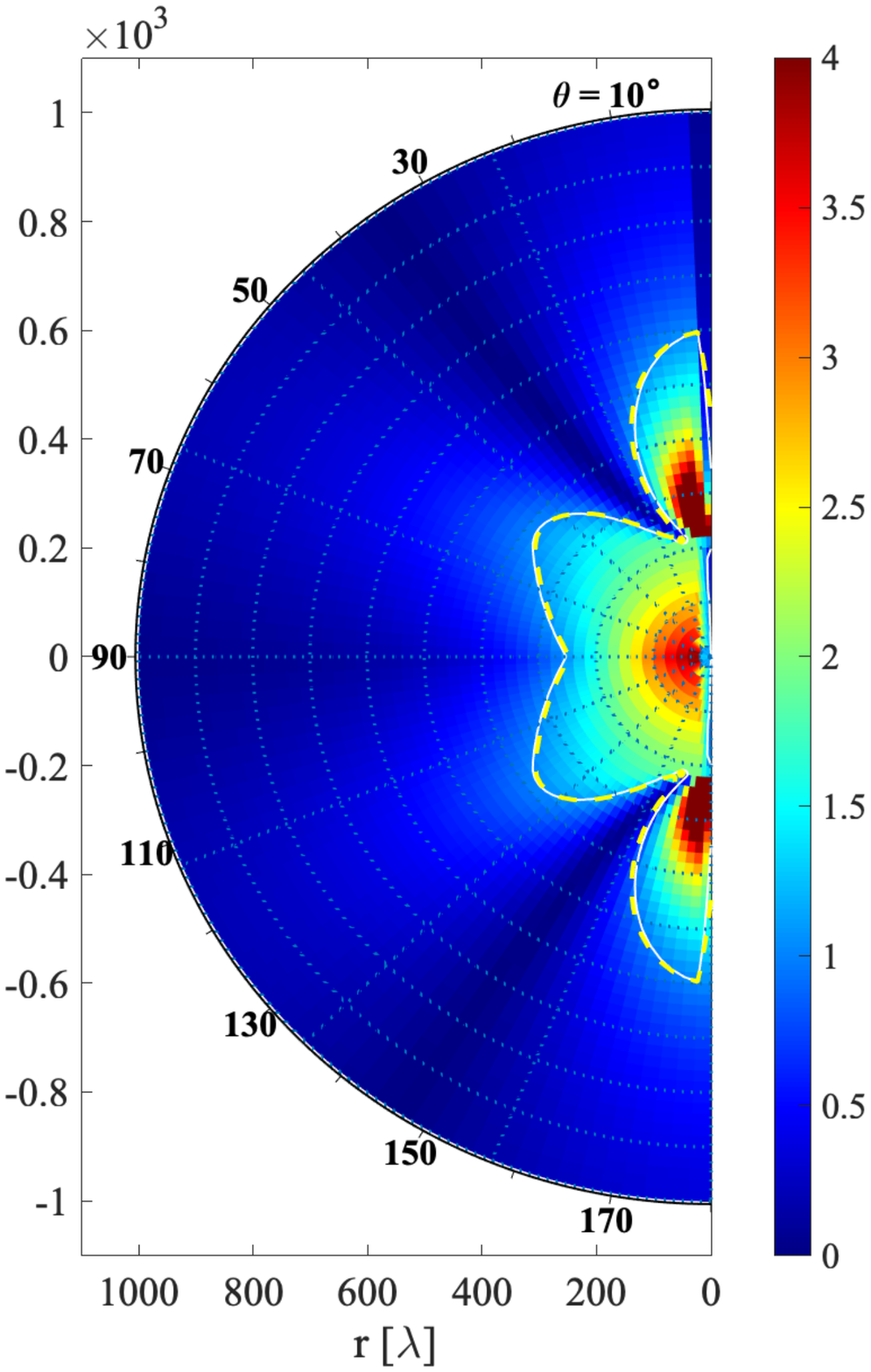} }
		\caption{Non-constant bandwidth spatial multiplexing region boundaries $R^{\mathrm nc,a}_{z}(\theta ;  L, \rho,\Delta K=1)$ and $R^{\mathrm nc,a}_{x}(\theta ;  L, \rho,\Delta K=1)$ (in dashed yellow lines) presented over the K number error color maps, for $r \in [0, 1000 \lambda]$ and $\theta \in [0, \pi]$, with $L = 400 \lambda$ and $\rho = 20\lambda$. The white contour lines represent the corresponding exact boundaries.} 
		\label{fig5}
	\end{figure}

\begin{figure}[!t]
		\centering 
		\includegraphics [width= .47\linewidth,trim= 95 115 110 110, clip]{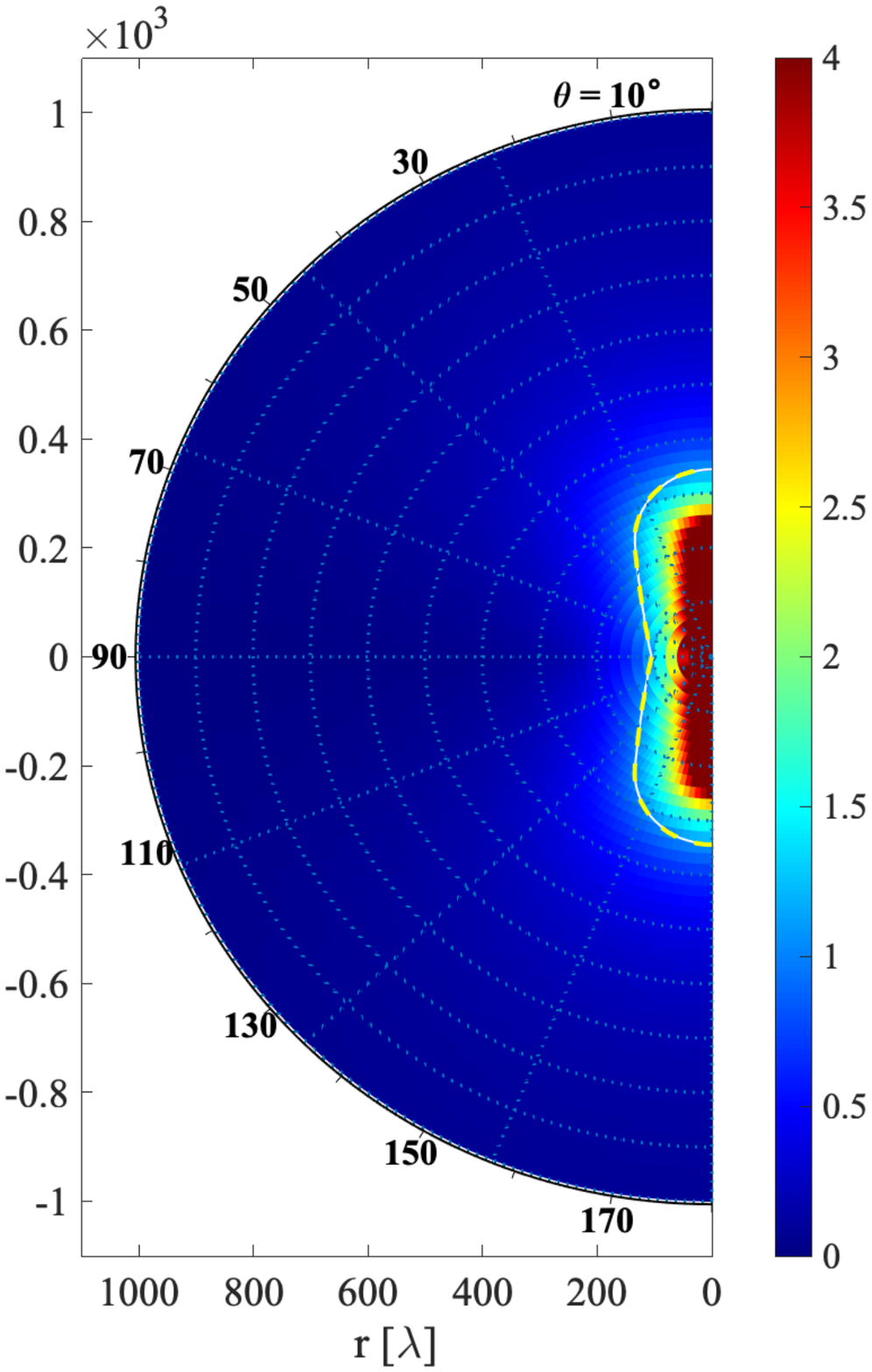}
		\caption{Spatial multiplexing region (non-constant bandwidth) boundary $R^{\mathrm a}_{y}(\theta ;  L, \rho,K_0=1)$ (in dashed yellow lines) presented over the K number color maps, for $r \in [0, 1000\lambda]$ and $\theta \in [0, \pi]$, with $L = 400\lambda$ and $\rho = 20\lambda$.  The white contour line represents the corresponding exact boundary.}
		\label{fig52}
	\end{figure}

Results for the five boundaries: $R^{\mathrm u}_{z}(\theta ;  L, \rho,K_0)$, $R^{\mathrm u}_{x}(\theta ;  L, \rho,K_0)$, $R_z^{\mathrm{nc,a}}(\theta; L,\rho,\Delta K)$, $R_x^{\mathrm{nc,a}}(\theta; L,\rho,\Delta K)$, and $R^{\mathrm a}_{y}(\theta ;  L, \rho,K_0)$, obtained by solving \eqref{eq:PW_zx}, \eqref{eq:SW_zx},  and \eqref{eq:SW_y}, are shown in Fig. \ref{fig4}, Fig. \ref{fig5}, and Fig. \ref{fig52},  represented by the yellow dashed lines. Their exact correspondences obtained directed based on the color maps of either the K number $K_{\hatbf{v}}(\overline{\mathbf{\Omega}})$ or the constant approximation error of the K number $K^{\mathrm u}_{\hatbf{v}}(\overline{\mathbf{\Omega}} ) - K_{\hatbf{v}}(\overline{\mathbf{\Omega}})$, where $\hatbf{v} \in \{\hat{\mathbf{e}}_{x}, \hat{\mathbf{e}}_{y} , \hat{\mathbf{e}}_{z} \}$, are represented using the white solid contour lines. A cutoff at $4$ is applied to all the color maps for better presentation. $K_{\hatbf{v}}(\overline{\mathbf{\Omega}})$ is computed by numerical integration following \eqref{eq:spatial_DoF_RCS}. Good agreements between the approximate boundaries and the exact boundaries can be observed in all these figures.

First of all, Fig. \ref{fig4}~(a) confirms the maximum distance threshold $R_0(L,\rho)$ for the spatial multiplexing region, seen by $\mathcal L_{z}$ at $\theta = \pi/2$.  It also clearly  shows the rapid decrease of  $R^{\mathrm u}_{z}(\theta ;  L, \rho)$ as $\mathcal L_z$ moves towards the endfire directions of $\mathcal L_s$, i.e., as $\theta \rightarrow 0$ or $\pi$.
On the other hand, Fig. \ref{fig4}~(b) shows that a relative direction near $\theta = \pi/4$ (or $\theta = 3 \pi/4$) is more conducive to the spatial DoF available in the LOS channel between $\mathcal L_s$ and $\mathcal L_x$, and $R^{\mathrm u}_{x}(\pi/4 ; L, \rho)$ is about half of $R_0(L,\rho)$. Moreover, by comparing  Fig.~\ref{fig4}~(a) and Fig.~\ref{fig4}~(b) we can see that $K_x$ is greater $K_z$ at the same distance when $\theta < \pi/4$ and  $\theta > 3 \pi/4$). 
It might be taken for granted that $\hat{\mathbf{e}}_{z}$ is the most advantageous orientation for a linear receiving array, but these results prove that $\hat{\mathbf{e}}_{x}$ is more conducive to create a large spatial bandwidth under the above $\theta$ conditions. Therefore, if the receiving array is enough far apart at a direction close to $\theta = \pi/2$, we can consider its projection in the $\hat{\mathbf{e}}_{z}$ direction only, because $\hat{\mathbf{e}}_{x}$ contributes very little in this case. However, if the receiving array is located in other directions, especially those that are closer to $\theta = \pi/4$ and $3 \pi/4$, this simplification can cause major errors. 
	
More interesting observations can be made from Fig.~\ref{fig5} and Fig.~\ref{fig52}. Firstly, as shown by Fig.~\ref{fig5}, a smaller distance $r$ does not necessarily mean severer changes in the local spatial bandwidth. In particular, under certain polar angle conditions, $\mathcal L_z$ may first enter and then leave the non-constant bandwidth spatial multiplexing region as $r$ decreases.  Moreover, for both $\mathcal L_{z}$  and $\mathcal L_{x}$, the polar angles that admit larger K numbers are also more favorable for the constant bandwidth approximation, as 
smaller $R_z^{\mathrm{nc,a}}(\theta; L,\rho)$ and $R_x^{\mathrm{nc,a}}(\theta; L,\rho)$ are seen.  The largest value over all $\theta \in (0, \pi)$ are about $600$~wavelengths for both $R_z^{\mathrm{nc,a}}(\theta; L,\rho)$ and $R_x^{\mathrm{nc,a}}(\theta; L,\rho)$, but appears at different polar angles. 
Generally speaking, the spatial locations closer to the two ends of $\mathcal L_s$ experience severer changes in the spatial bandwidth for both $\mathcal L_{z}$ and $\mathcal L_{x}$. For $\mathcal L_{y}$, however, drastic changes in the spatial bandwidth are experienced at any locations close to $\mathcal L_s$, as shown by Fig. \ref{fig52}. Finally, we note that although the $\hat{\mathbf{e}}_{y}$ direction can contribute to the achievable spatial DoF, the order of its contribution is much smaller than the other two directions.

	\begin{figure*}[t]
		\centering
		\subfigure[${K}^{a}_{z}(\overline{\mathbf{\Omega}} ) - K_{z}(\overline{\mathbf{\Omega}} ) $ ]{\includegraphics [width= .24\linewidth,trim= 95 115 95 110, clip]{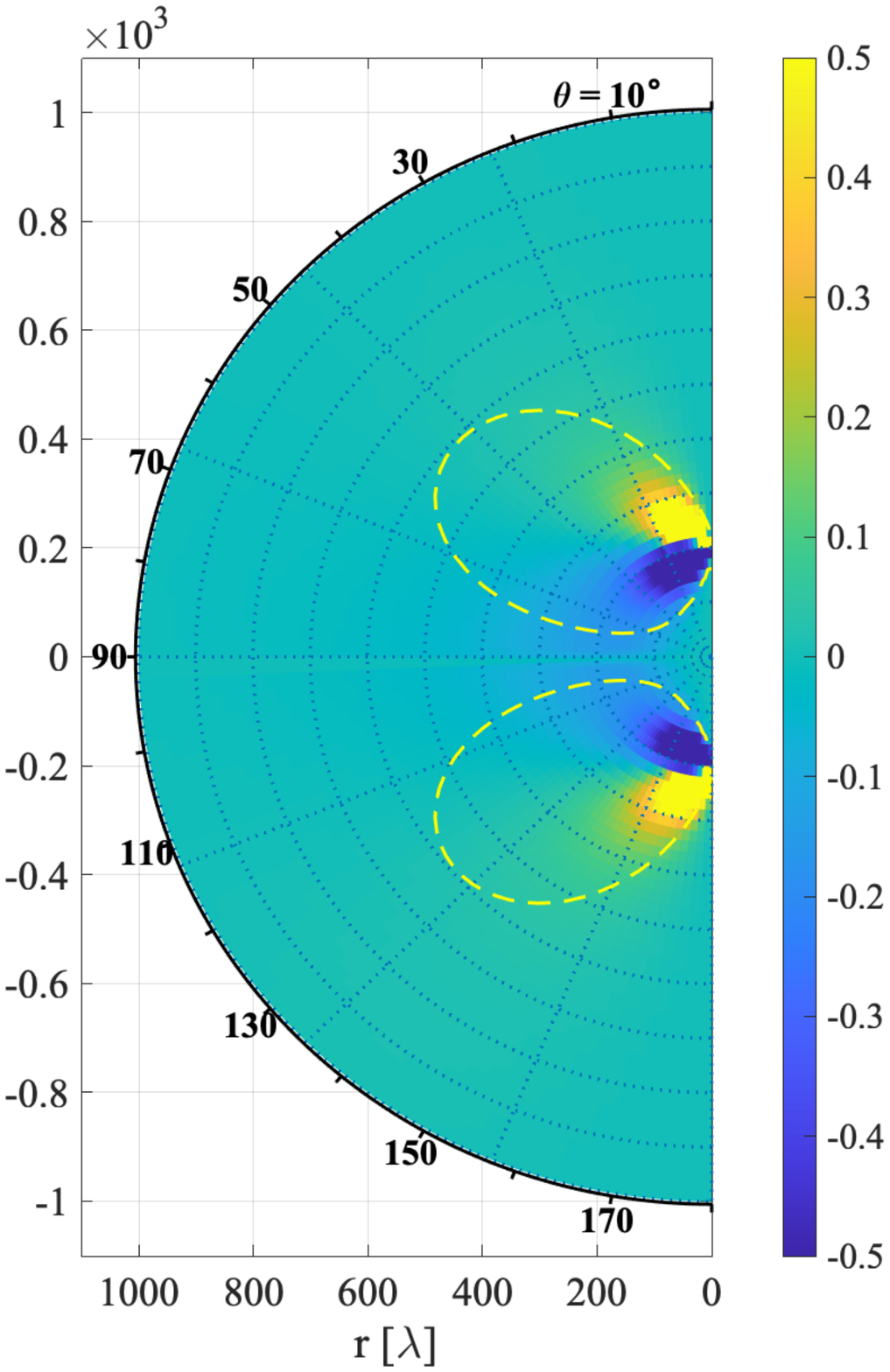} }
		\hspace{1cm}
		\subfigure[${K}^{a}_{x}(\overline{\mathbf{\Omega}} ) - K_{x}(\overline{\mathbf{\Omega}} ) $]{\includegraphics [width= .24\linewidth,trim= 95 115 95 110, clip]{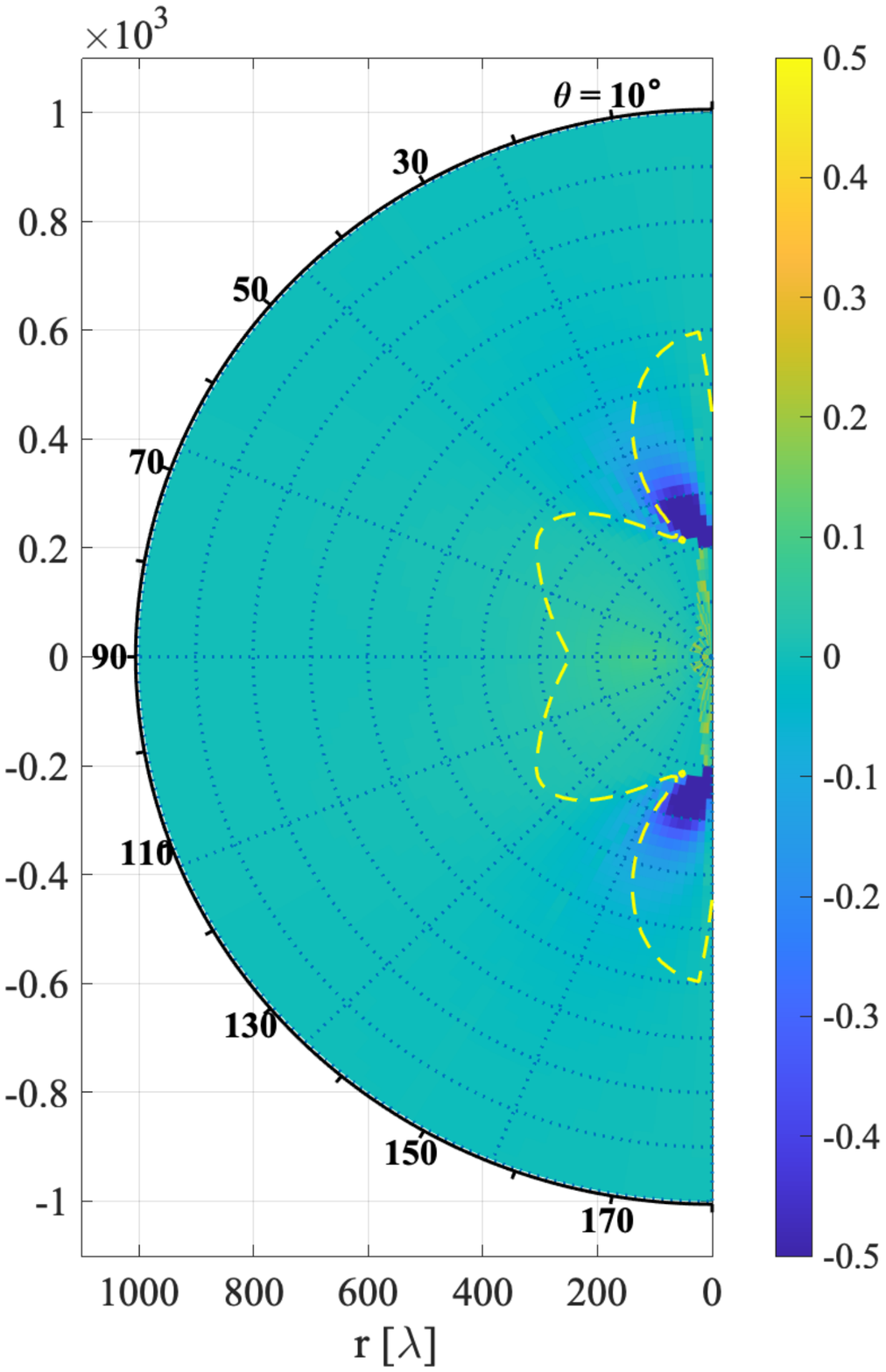} }
		\hspace{1cm}
		\subfigure[${K}^{a}_{y}(\overline{\mathbf{\Omega}} ) - K_{y}(\overline{\mathbf{\Omega}} ) $]{\includegraphics [width=.24\linewidth,trim= 95 115 95 110, clip]{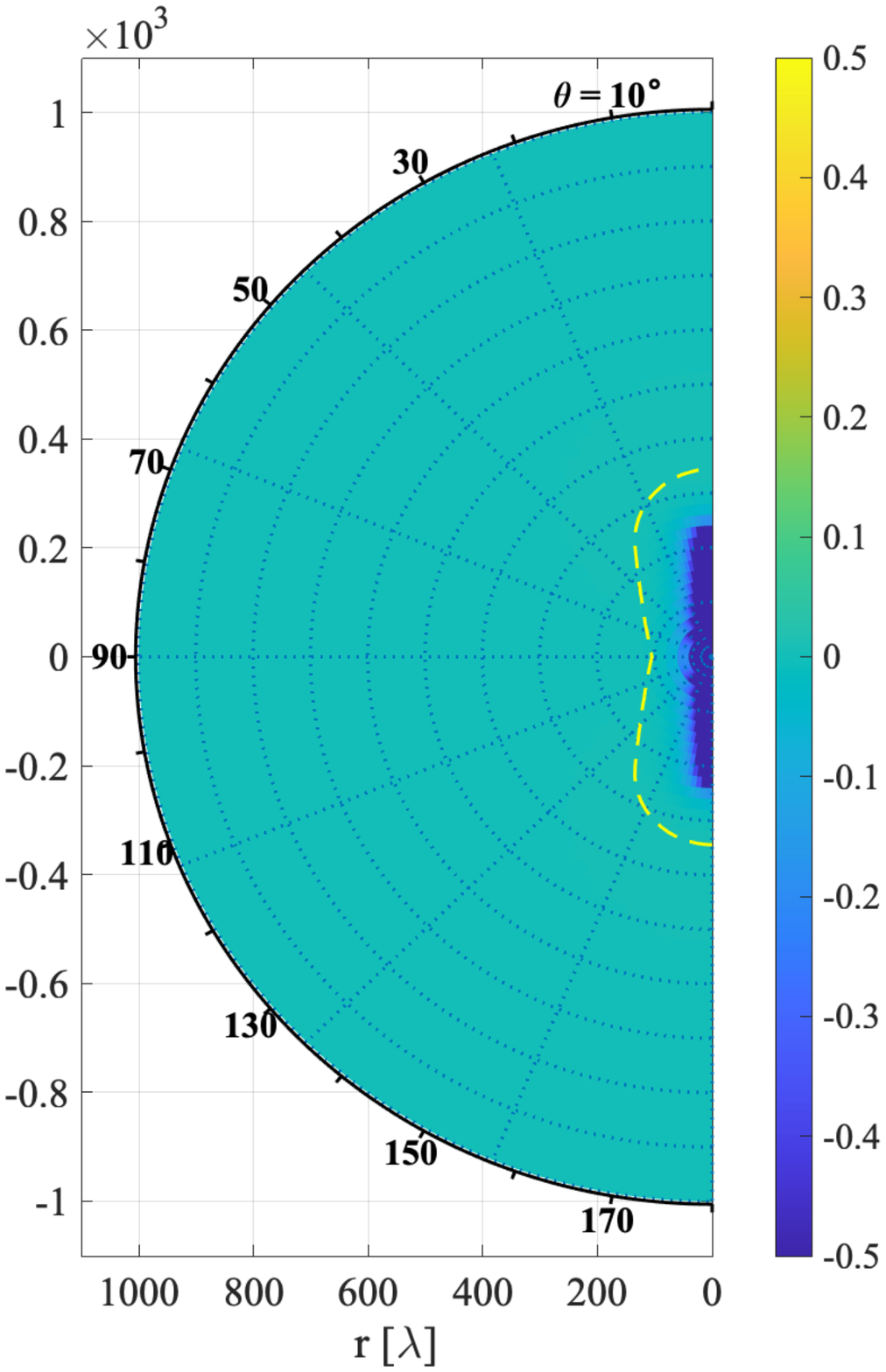} }
		\caption{K number calculation errors caused by the linear approximations \eqref{eq:DoF_lin_apx_z}, \eqref{eq:DoF_lin_apx_x} and \eqref{eq:DoF_lin_apx_y}, for $r \in [0, 1000 \lambda]$ and $\theta \in [0, \pi]$, with $L = 400 \lambda$ and $\rho = 20 \lambda$. }
		\label{fig6}
	\end{figure*}

In Fig. \ref{fig6}, the calculation errors in the K number caused by the proposed linear approximation formulas  \eqref{eq:DoF_lin_apx_z}, \eqref{eq:DoF_lin_apx_x} and \eqref{eq:DoF_lin_apx_y} are presented. As the results shows, the errors are very small even in most part of the non-constant bandwidth spatial multiplexing regions for all three receiving directions. In particular, for the $\hat{\mathbf{e}}_{z}$ and $\hat{\mathbf{e}}_{x}$ directions, large errors only appear at the locations that are very close to the two ends of $\mathcal L_s$, which ensures the validity of these linear approximations in many practical scenarios. 
	
To conclude, under this particular simulation setup, the sizes of the non-constant-bandwidth spatial multiplexing regions are in the same scale as $\mathcal L_s$, but the shapes are irregular and very different for the three receiving directions. For many practical application scenarios, sufficient separation between the source and receiving arrays can be ensured, for instance, by taking advantage of the height differences between the network infrastructure and the user terminals. As a result, a constant spatial bandwidth approximation can be used without the risk of large approximation errors. Unless the array assembly is designed for very short distance communication and more fine-grained  analysis is demanded, the proposed linear approximation formulas \eqref{eq:DoF_lin_apx_z}, \eqref{eq:DoF_lin_apx_x} and \eqref{eq:DoF_lin_apx_y} can be applied. 
	
	\begin{figure*}[!t]
		\centering 
		\subfigure[]{\includegraphics [width= .4\linewidth,trim= 30 225 40 240, clip]{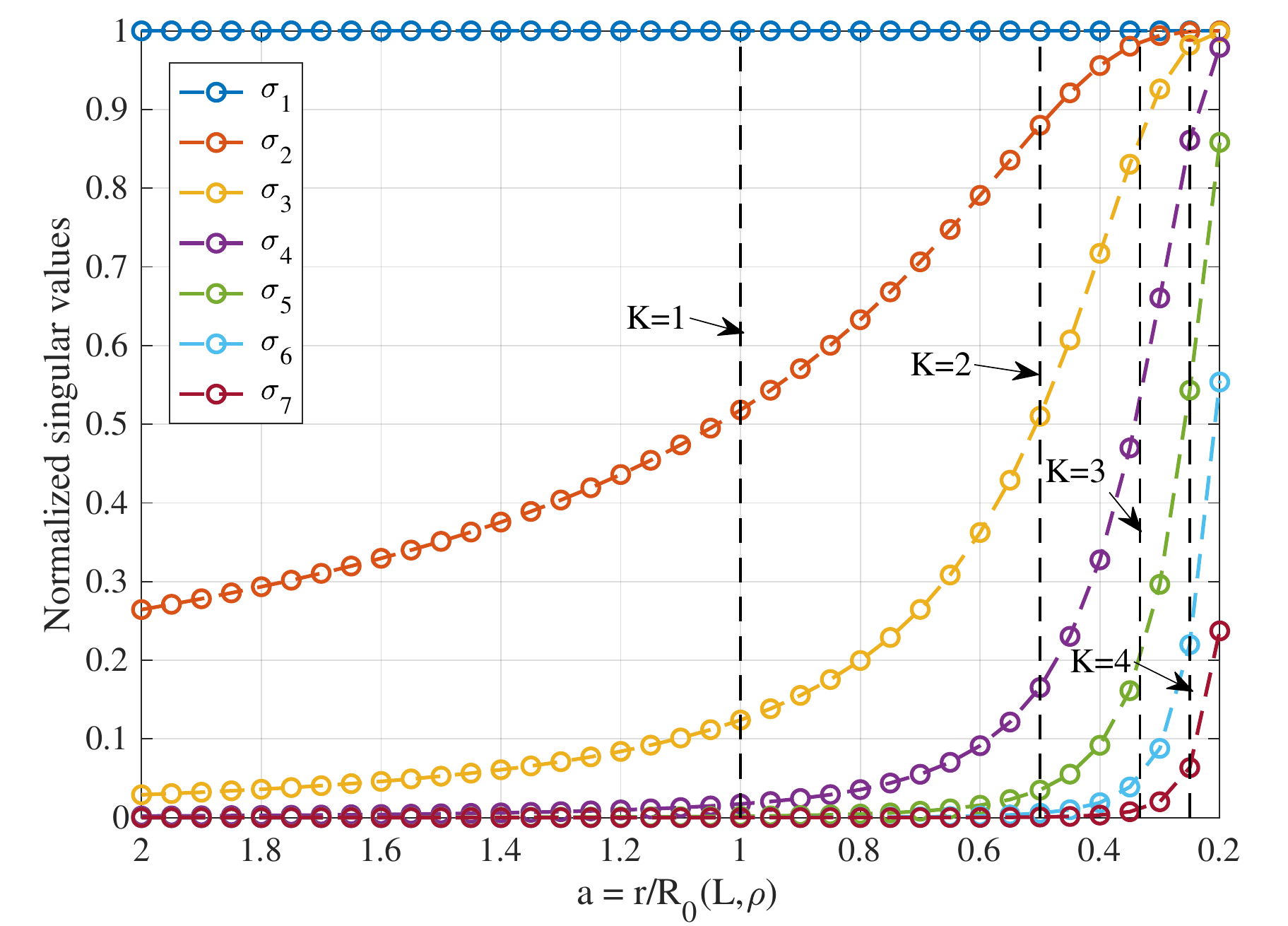} }
		\hspace{1cm}
		\subfigure[ ]{\includegraphics [width= .4\linewidth,trim= 30 225 40 240, clip]{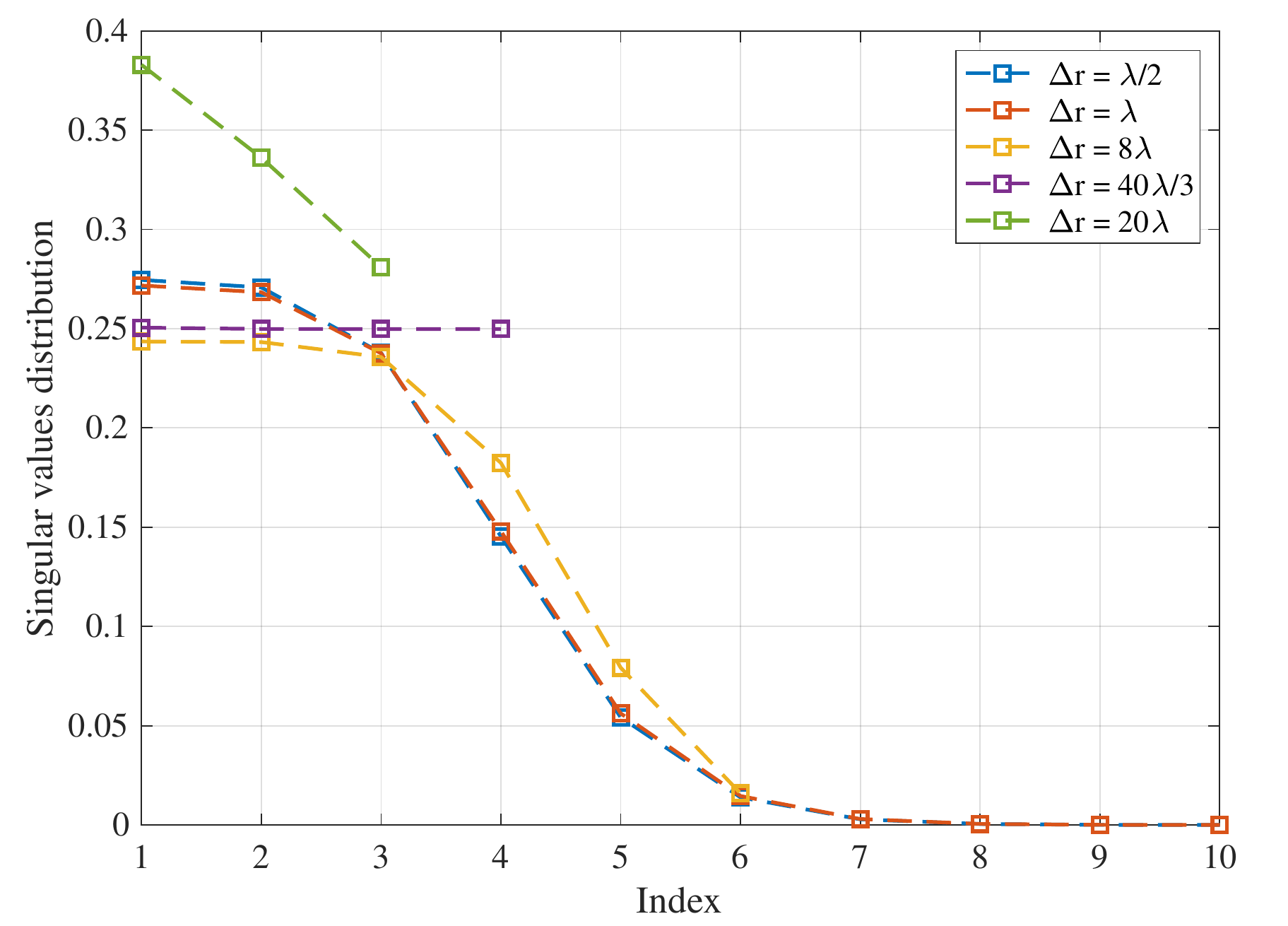} }
		\caption{Singular value performance of the discrete LOS channel matrix between $\mathcal L_r$ and $\mathcal L_s$, with uniform antenna spacing, $\theta = \pi/2$. (a) Comparison of the normalized singular values (in decreasing order, showing only the first $7$) over the decreasing distance $r = a R_0(L,\rho)$, $\Delta_s =\Delta_r = \lambda/2$. 
		(b) The distribution of singular values (in decreasing order, showing up to the first $10$), obtained at a distance achieving $K_z =3$, with $\Delta_s =\lambda/2$ and different $\Delta_r$.}
		\label{fig:singular}
	\end{figure*}
	
Finally, by examining the singular value distribution of the discretized LOS channel between $\mathcal L_s$ and $\mathcal L_z$ with $\theta = \pi/2$, we validate the appropriateness of using small $K_0$ for spatial multiplexing region demarcation and demonstrate the potential benefits  of the K number analysis. As shown by Fig.~\ref{fig5}(a), constant spatial bandwidth approximation is good in this setting for any choice of $r$. Therefore, uniform antenna spacing is applied to both arrays. Denoting $\Delta_s$ and $\Delta_r$ to be the antenna spacing, the number of antennas deployed by $\mathcal L_s$ and $\mathcal L_z$ are thus given by $N_t = 1+L/\Delta_s$ and $N_r = 1+2\rho/\Delta_r$, respectively. A uni-polarization situation is assumed for simplicity. Then following \eqref{eq:dyadic_Green_farfield}, the $(i,j)$-th element of the  $N_r\times N_t$ channel matrix $\mathbf H$ is modeled by $r^{-1}_{i,j} \exp(j2\pi r_{i,j})$, where  $r_{i,j}$ is the actual distance between the $i$-th antenna on $\mathcal L_z$ and the $j$-th antenna on $\mathcal L_s$. The singular values are obtained by performing SVD to $\mathbf H$. 

The distribution of the singular values (sorted in decreasing order), obtained with $\Delta_s = \Delta_r = \lambda/2$ ($N_t =801$, $N_r =81$) under different choices of distance $r$, are shown in Fig. \ref{fig:singular}(a) against $a \triangleq \frac{r}{R_0(L,\rho)}$. 	To eliminate the influence of different power loss levels at different distances, the singular values are normalized using the largest one for each channel matrix. The distances achieving $K_z = 1$, $2$, $3$ and $4$, calculated using \eqref{eq:Rz_K0}, are also marked. When $a =1$, $r \approx R_0(L,\rho)$, $K \approx 1$, $\sigma_2$ is slightly larger than $0.5$, indicating the existence of a subchannel of power gain just exceeds a quarter of the best one. It may be plausible to adopt $0.3$ as the threshold for determining the usability of a subchannel, as it represents a power loss of approximately $10$ dB relative to the best one. With this threshold, we see that with $a = 0.5$, $K \approx 2$, $3$ usable subchannels exist; with $a = 0.4$, $K\approx 2.5$, $\sigma_4$ just reaches the threshold; and with $a = 0.3$, $K\approx 3.5$, $\sigma_5$ falls slightly below.  Therefore, the number of significant eigenvalues of the channel are closely related to the K number even when it is small, and it is practically meaningful to use a small $K_0$ to determine the spatial multiplexing regions following Definition \ref{Def:spatial_multiplexing_region}. 
	
To show the potential benefit of the K number analysis, we compare in Fig. \ref{fig:singular}(b) the singular value distribution obtained under five different choices of $\Delta_r$ (all divide $2\rho$), when $r$ is fixed such that $K_z =3$ and $\Delta_s = \lambda/2$ stays unchanged. To be precise, we plot $\sigma_1/A, \sigma_2/A, \ldots, \sigma_{N_r}/A$, where $A = \sum_{i=1}^{N_r} \sigma_i$, for different $\Delta_r$. In particular, by choosing $\Delta_r = 40\lambda/3$, which corresponds to the Nyquist sampling interval given a constant bandwidth $K_z/2\rho$ \cite{migliore2006role}, we expect the MIMO channel to have $4$ equally good singular values ($N_r =4$ in this case). Choosing $\Delta_r = \lambda/2$ and $\lambda$ corresponds to two oversampling situations with small spacing, 	$\Delta_r = 8\lambda$ to an oversampling situation with relative large spacing, and $\Delta_r = 20\lambda$ to an undersampling situation. Firstly, Fig.~\ref{fig:singular}(b) clearly supports our expectation of the optimal sampling  $\Delta_r = 40\lambda/3$, as all $4$ singular values are practically the same. On the other hand, both oversampling and undersampling lead to uneven singular value distribution.  Owing to the inherent achievable spatial DoF limitation, for all three oversampling cases, the singular value transition window starts at the fourth and ends at the seventh. Moreover, the small spacing results in more severe unevenness in the first four. 

\section{LOS Communications with Rotatable Linear Receiving Array}
\label{sec:7}
	
	\begin{figure*}[t]
		\centering 
		\subfigure[Vertical placement] {\includegraphics [height= 4.8cm] {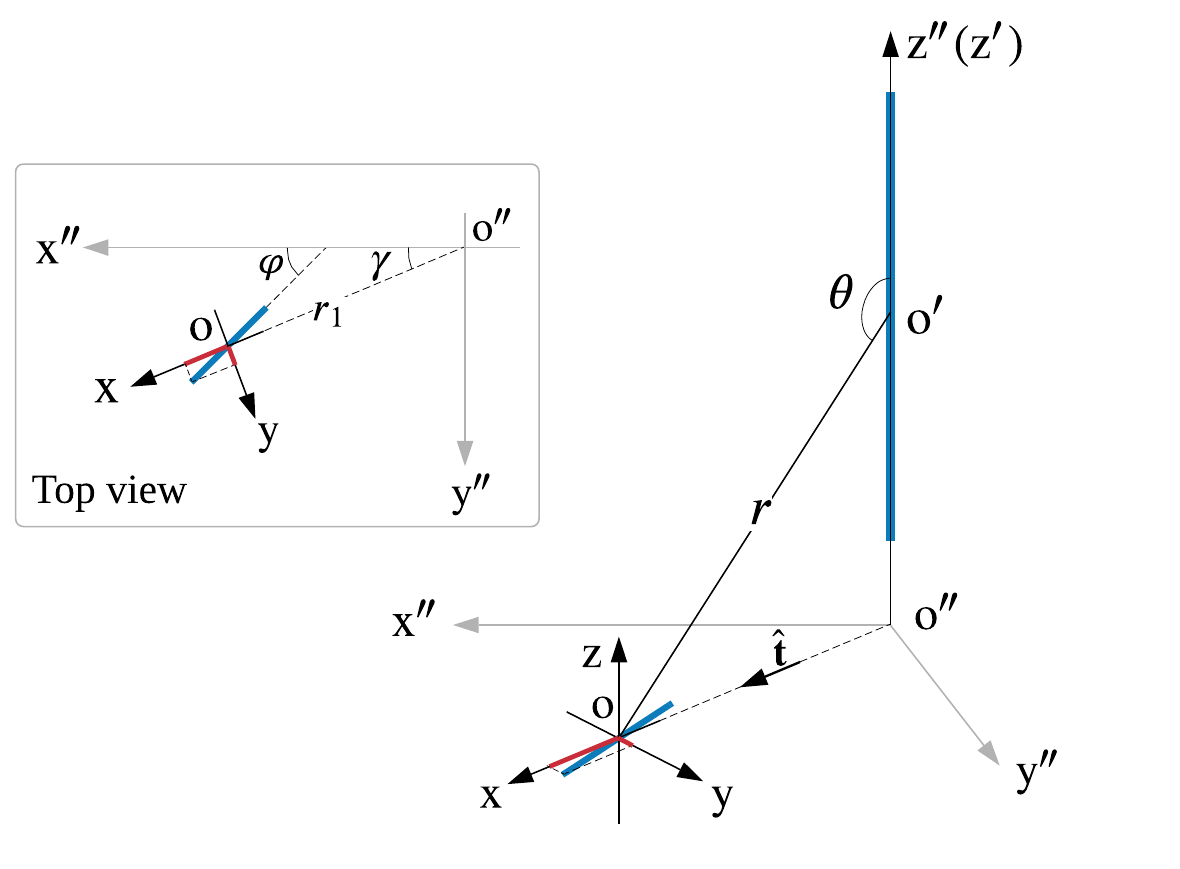} } \hspace{2em}
		\subfigure[Horizontal placement]  {\includegraphics [height= 4.8cm] {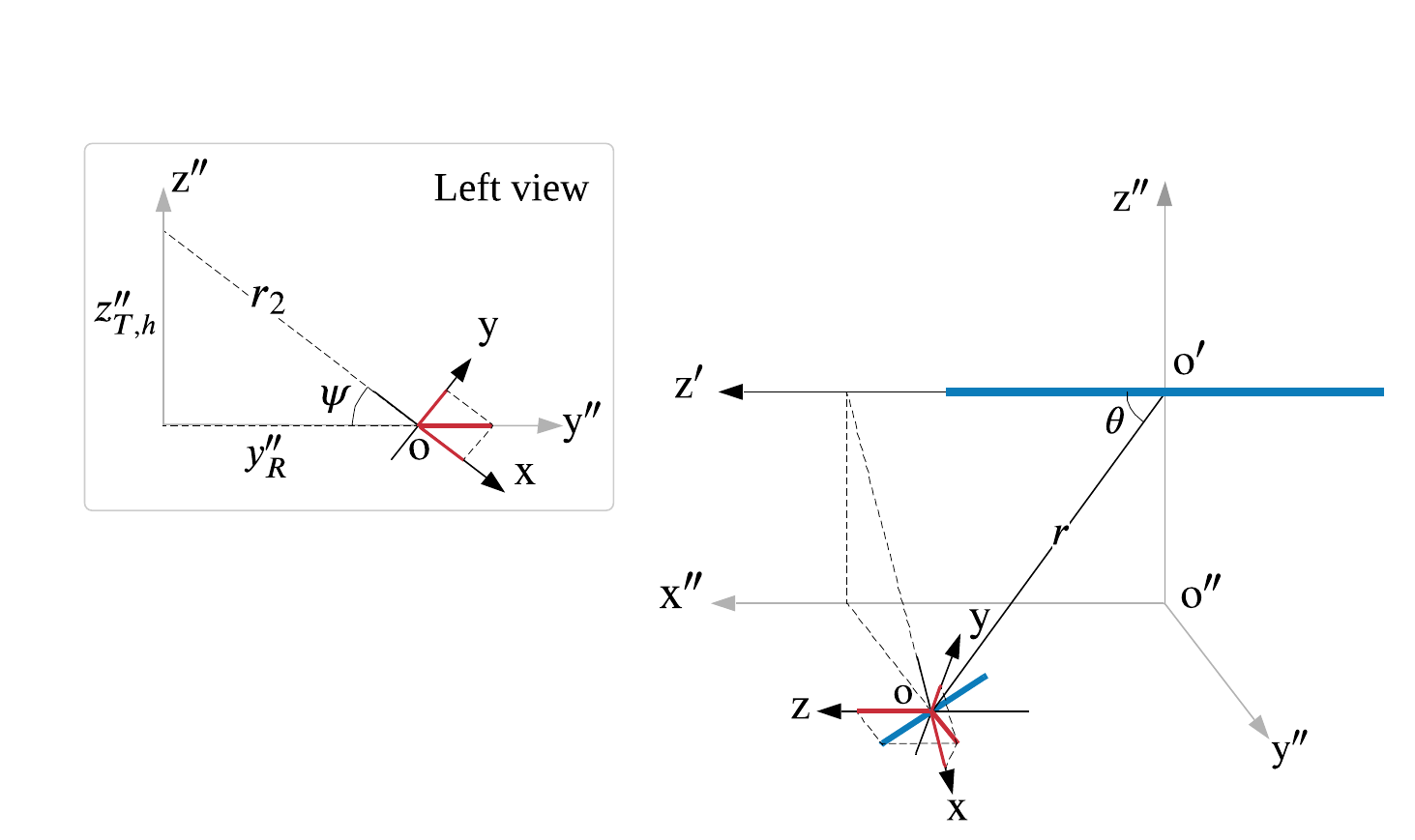} }
		\caption{Scenarios for the LOS channel study between an elevated source LSAA and a linear receiving array that can freely rotate and move on the ground. }
		\label{fig8}
	\end{figure*}
	
To demonstrate the potential application of the proposed analytical framework, the derived spatial bandwidth results, and the understandings we have gained through the previous sections, we study the achievable spatial DoF performance of the LOS channel between a source LSAA $\mathcal L_s$  and a small linear receiving array $\mathcal L_r$  in two simple scenarios, shown in Fig. \ref{fig8}.  In particular, $\mathcal L_s$ is deployed at an elevated position, aligned vertically or horizontally with the ``ground plane'', on which $\mathcal L_r$ is located and can rotate freely. 	A world coordinate system $(\mathrm{x}'', \mathrm{y}'', \mathrm{z}'')$ with origin  $\mathrm{o}''$ set on the ground plane is defined and used as the default coordinate system. The center of $\mathcal L_s$ is given by $\mathrm{o}' = (0, 0, z''_{S,v})$ for the vertical placement scenario and $\mathrm{o}' = (0, 0, z''_{S,h})$ for the horizontal placement scenario. The center of  $\mathcal L_r$ is given by $\mathrm{o} = (x''_R, y''_R, 0)$, where $y''_R \geq 0$ is required to focus the discussion on the $+\mathrm{y}''$ half-plane.  The situation on the other half-plane is simply a mirror image.  The lengths of the two arrays are denoted using $L$ and $L_r$, respectively.  $L = 400 \lambda$, $L_r  = 40 \lambda$, $z''_{S,v} =  400 \lambda$, and $z''_{S,h} = 200 \lambda$ are set for the study. 
	
Given  $\mathrm{o}'$ and  $\mathrm{o}$, their distance $r$ is trivial to obtain. The unit directional vector of $\mathcal L_r$ is denoted using $\hatbf{v}$,  and  the unit directional vector of the $\mathrm{o}''$-$\mathrm{o}$ connecting line  is denoted as $\hatbf{t} =  \big({x''_R}/{r_1}, {y''_R}/{r_1}, 0 \big)$, where  $r_1 =  \sqrt{ (x''_R)^2 + (y''_R)^2}$. Moreover, as shown in the {top-view} diagram in Fig.~\ref{fig8}(a), the \textit{orientation angle} of $\mathcal L_r$ is defined as $\varphi \triangleq \arccos ( \langle \hatbf{v} , \hat{\mathbf{e}}_{x''} \rangle )$ and thus, $ \varphi \in [0, \pi]$. The angle between the $\mathrm{o}''$-$\mathrm{o}$ line and the $\mathrm{x}''$-axis is given by  $\gamma({\mathrm{o}})  \triangleq \arccos( \langle \hatbf{t} , \hat{\mathbf{e}}_{x''} \rangle ) = \arccos( {x''_R}/{r_1} )$.  
	
\subsection{Vertical Tx Array Placement Scenario}
\label{sec:7a}
	
This scenario is shown in  Fig.~\ref{fig8}(a). Following Section~\ref{sec:4a},  an $\mathrm{o}'$-origin coordinate system with the $\mathrm{z}'$-axis coincides with the $\mathrm{z}''$-axis, and an $\mathrm{o}$-origin  receiving  coordinate system with the three coordinates denoted using $(\mathrm{x}, \mathrm{y}, \mathrm{z})$ are defined. In particular, the $\mathrm{z}$-axis is parallel with the $\mathrm{z}''$-axis and the $\mathrm{x}$-axis is in the same direction as the $\mathrm{o}''$-$\mathrm{o}$ line. Hence, $\hat{\mathbf{e}}_{x} \equiv \hatbf{t}$,  	and the polar angle $\theta$ is  given by $\theta = \arctan^* \big( {r_1 }/{z''_{S,v}} \big)$, where the  $\arctan^*(\cdot)$  function is defined as 
\begin{equation*}
	\arctan^*(x) \triangleq    \pi  \cdot \mathrm{sign^-}(x)  + \arctan(x) 
\end{equation*}
with the operator $\mathrm{sign^-}(x)$ returns $1$ if $x<0$ and $0$ if $x\geq 0$.

\begin{figure*}[!t]
		\centering
		\subfigure[$\varphi = 0$] {\includegraphics [width=.41\linewidth,trim=0 275 30 275, clip]{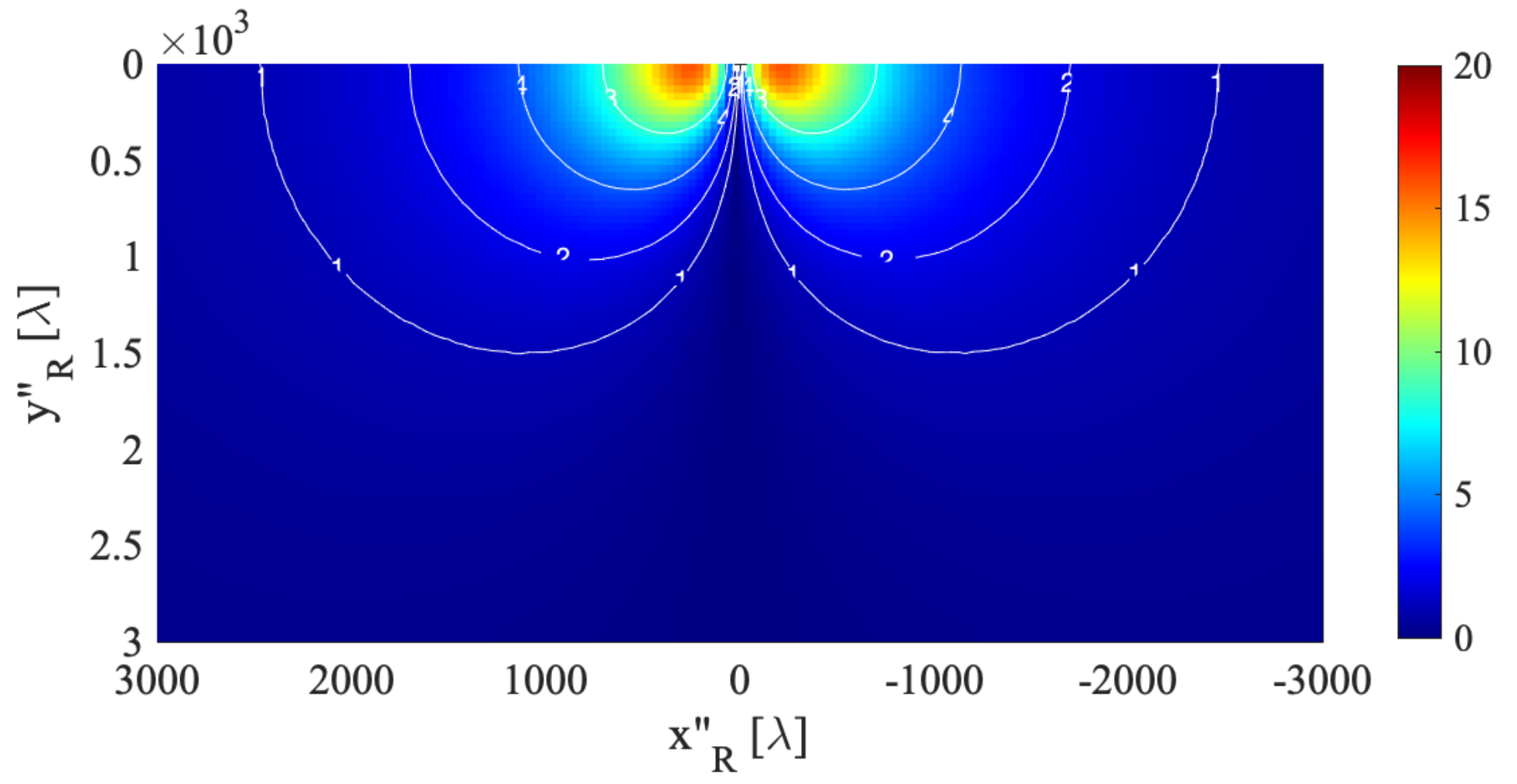} }
		\hspace{0.8cm}
		\subfigure[$\varphi = \pi/4$]  {\includegraphics[width=.41\linewidth,trim=0 275 30 275, clip]{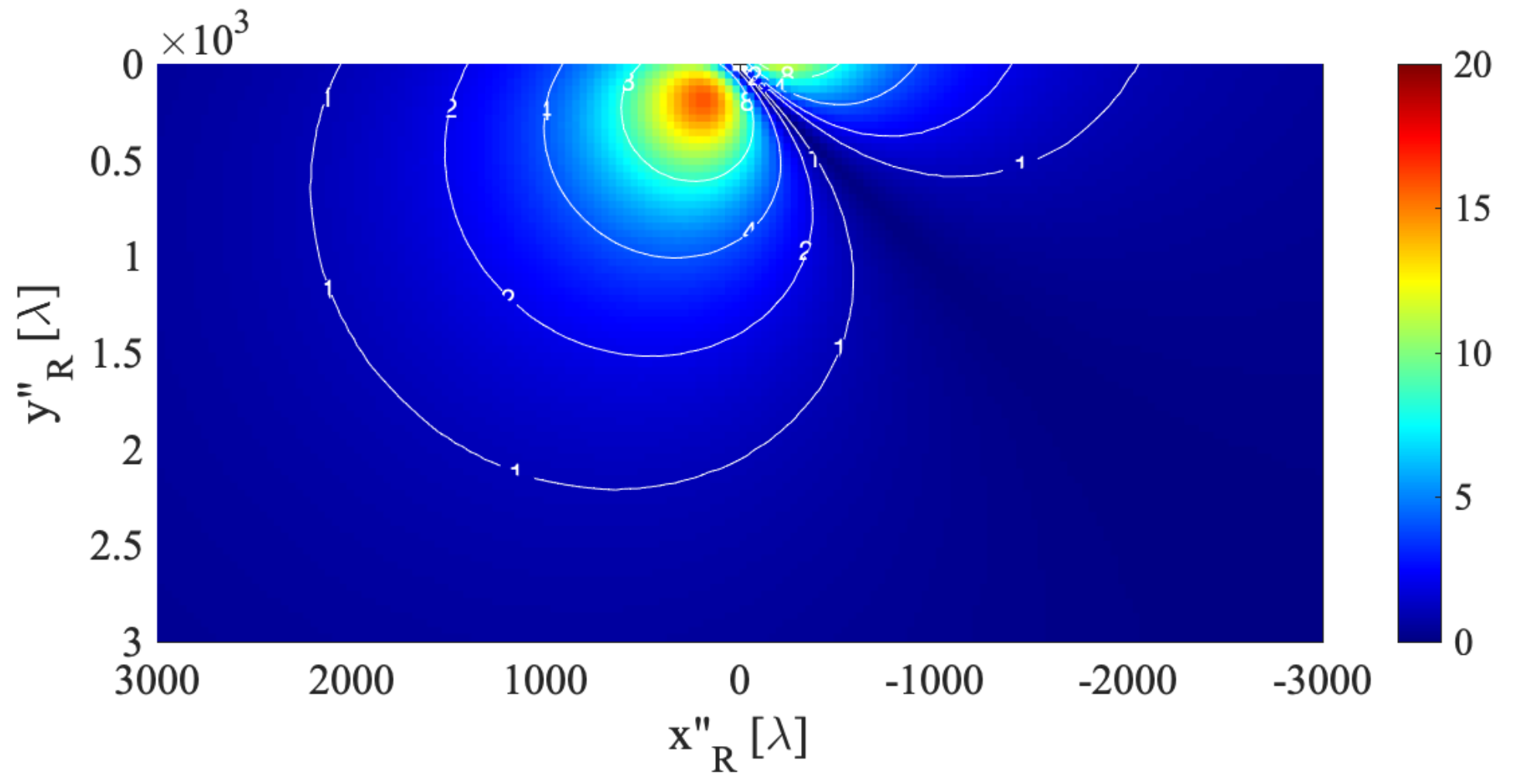} }
		\subfigure[$\varphi = \pi/2$] {\includegraphics [width=.41\linewidth,trim=0 275 30 275, clip]{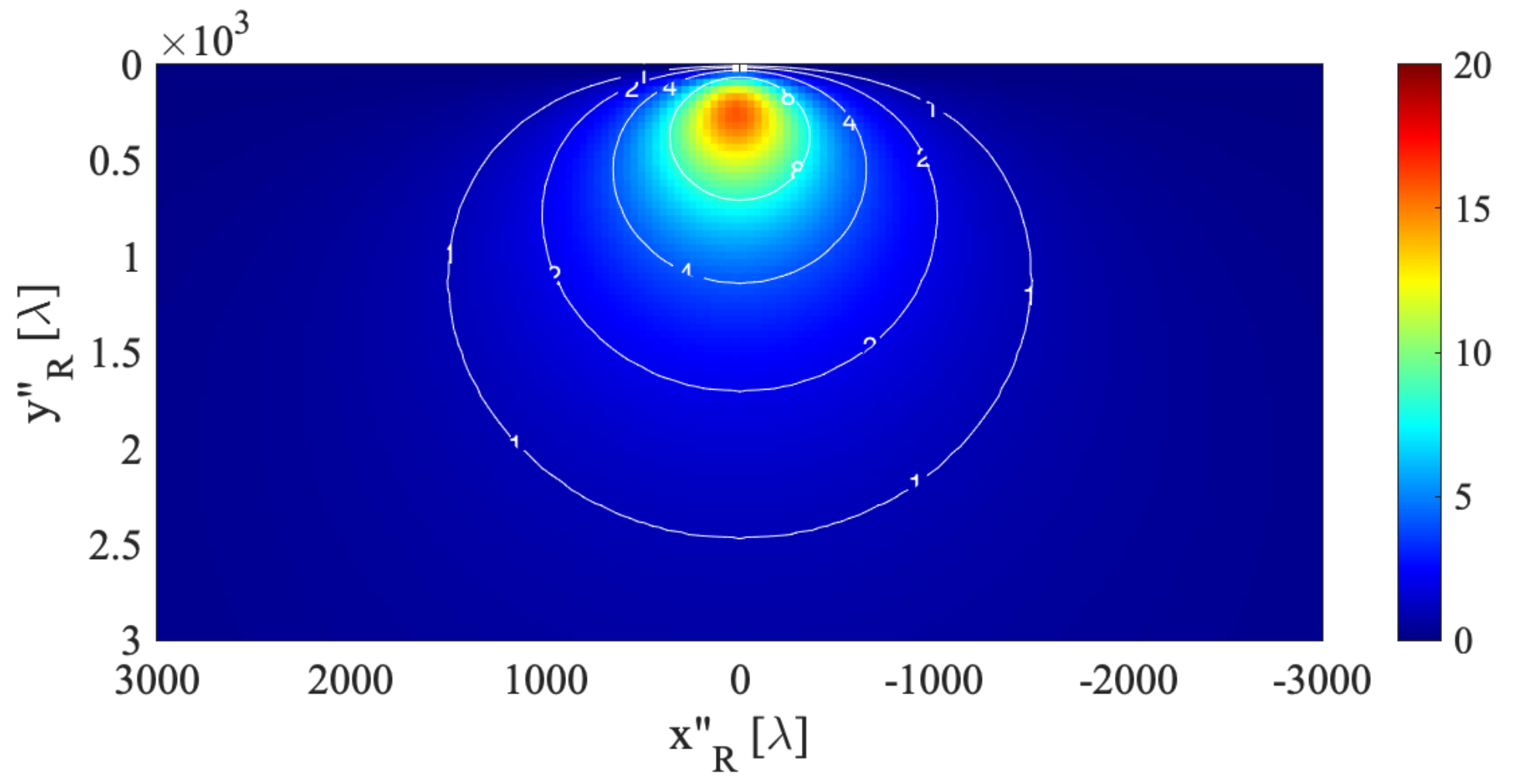} }
		\hspace{0.8cm}
		\subfigure[$ \varphi  = \gamma({\mathrm{o}}) $ ]  {\includegraphics[width=.41\linewidth,trim=0 275 30 275, clip]{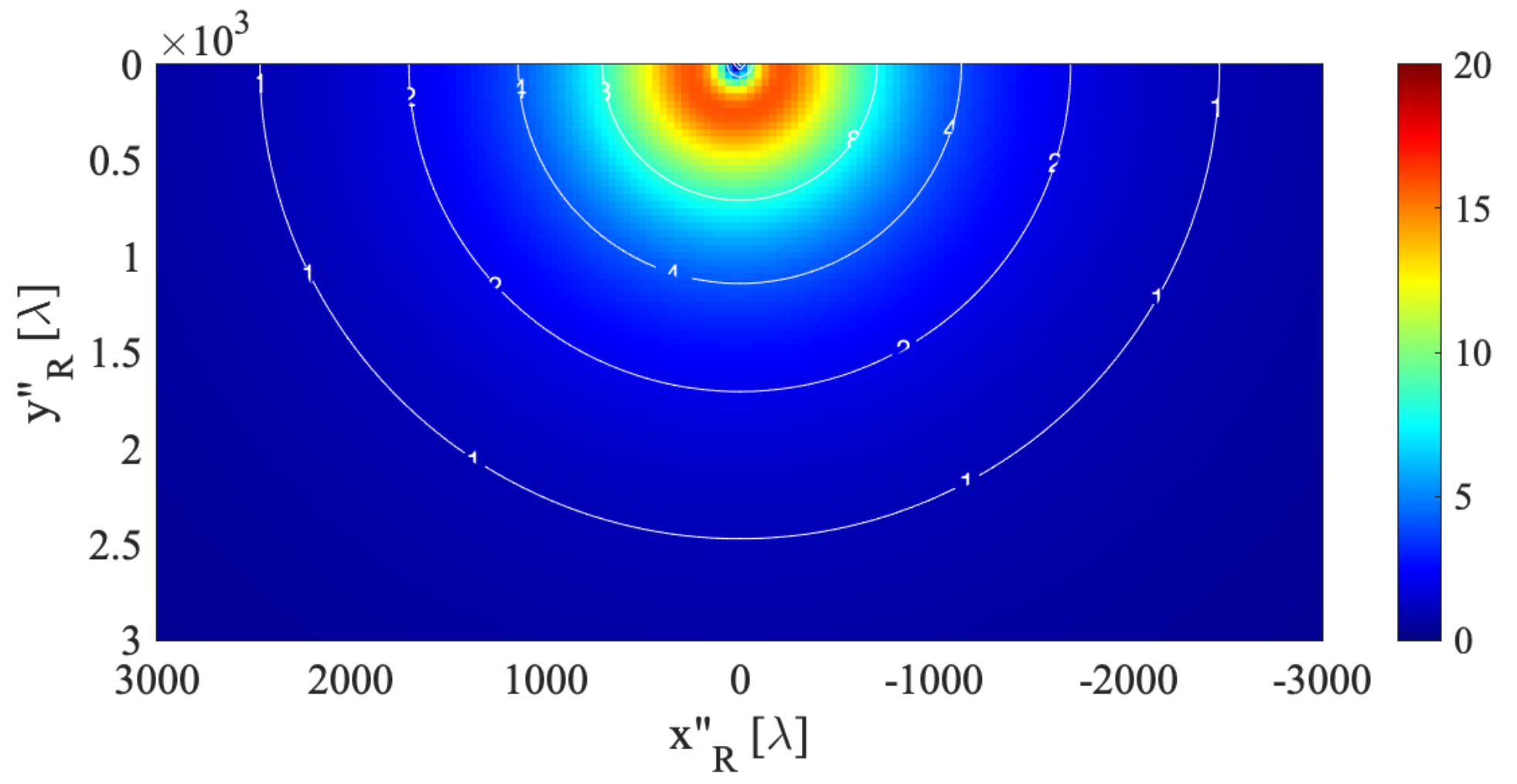} }	
		\caption{The K numbers achieved by $\mathcal L_r$ with different orientation angles on the $+\mathrm{y}''$-half ground plane in the vertical LSAA placement scenario, with $L = 400 \lambda$,  $\rho = 20 \lambda$, and  $z''_{S,v} = 400 \lambda$.  Contour lines at $K = 1$, $2$, and $4$ are shown.}
		\label{fig:DoF_vTx}
\end{figure*}
	
The projection of $\mathcal L_r$ is zero in the $ \hat{\mathbf{e}}_{z}$ direction, preventing it from making any contribution to spatial DoF. The projections in the other two directions are given by $L^v_{R,x}  =  L_r |\cos(\varphi - \gamma({\mathrm{o}}) )|$ and  $L^v_{R,y}  = L_r \left|\sin(\varphi - \gamma({\mathrm{o}}) ) \right| $, respectively.  Based on Fig. \ref{fig52}, we know that the contribution from the $\hat{\mathbf{e}}_{y}$ direction is negligible unless $(z''_{S,v} - L/2 )< 150$ (in this special setup) and $\mathrm{o}$ is very close to $\mathrm{o}''$. Therefore, we approximately consider $\hat{\mathbf{e}}_{x}$ as the only contributing direction and suggest the use of $ \varphi  = \gamma({\mathrm{o}}) $ to  maximize $L^v_{R,x}$, and thus, the resulting K number, if the orientation of $\mathcal L_r$ can be controlled. Moreover, it is easy to infer that by choosing $ \varphi  = \gamma({\mathrm{o}}) $, the K number will remain the same when $\mathcal L_r$  moves on a circle centered at $\mathrm{o}''$. Moreover, based on Fig.~\ref{fig4}(b), it can also be inferred that the best  K number performance can be expected when $r_1$ is comparable with $z''_{S,v} $, and thus $\theta$ is close to  $3\pi/4$.  Meanwhile, if $\mathcal L_r$  is very close to $\mathrm{o}''$, the  K number performance will be poor.
	
Fig. \ref{fig:DoF_vTx} presents the color maps of the K number  obtained numerically following \eqref{eq:spatial_DoF_RCS} over the $+\mathrm{y}''$-half ground plane, for three fixed orientations $\varphi = 0$, $\pi/4$ and $\pi/2$ and the location-dependent orientation $ \varphi  = \gamma({\mathrm{o}})$.  The results are consistent with the discussion above.  In particular, from Fig.~\ref{fig:DoF_vTx}(a)--Fig.~\ref{fig:DoF_vTx}(c) one can see that the worst performance appears at the positions causing $|\varphi -  \gamma({\mathrm{o}}) | = \pi/2$ and thus $L^v_{R,x}  =  0$. Therefore, without orientation  control, certain location and orientation conditions necessarily result in poor K number performance. On the other hand, with the simple orientation control  $ \varphi  = \gamma({\mathrm{o}})$, consistent K number performance is assured when  $\mathcal L_r$ is located at same radial distance to $\mathrm{o}''$. 
	
\subsection{Horizontal Tx Array Placement Scenario}
\label{sec:7b}
	
	\begin{figure*}[t]
		\centering
		\subfigure[$\varphi = 0$] {\includegraphics [width=.41\linewidth, trim=10 210 30 220, clip]{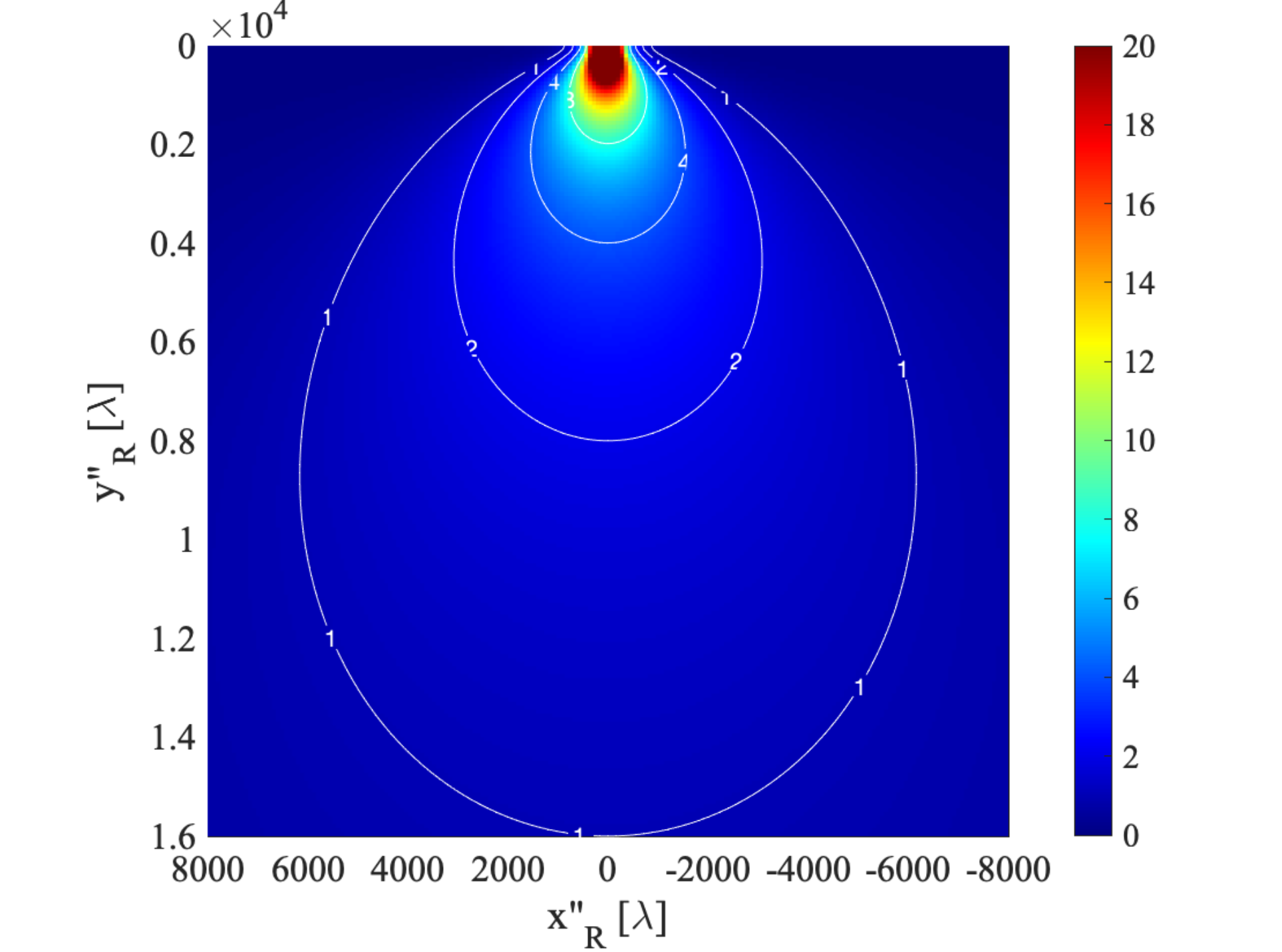} }\hspace{0.8cm}
		\subfigure[$\varphi = \pi/4$]  {\includegraphics[width=.41\linewidth, trim=10 210 30 220, clip]{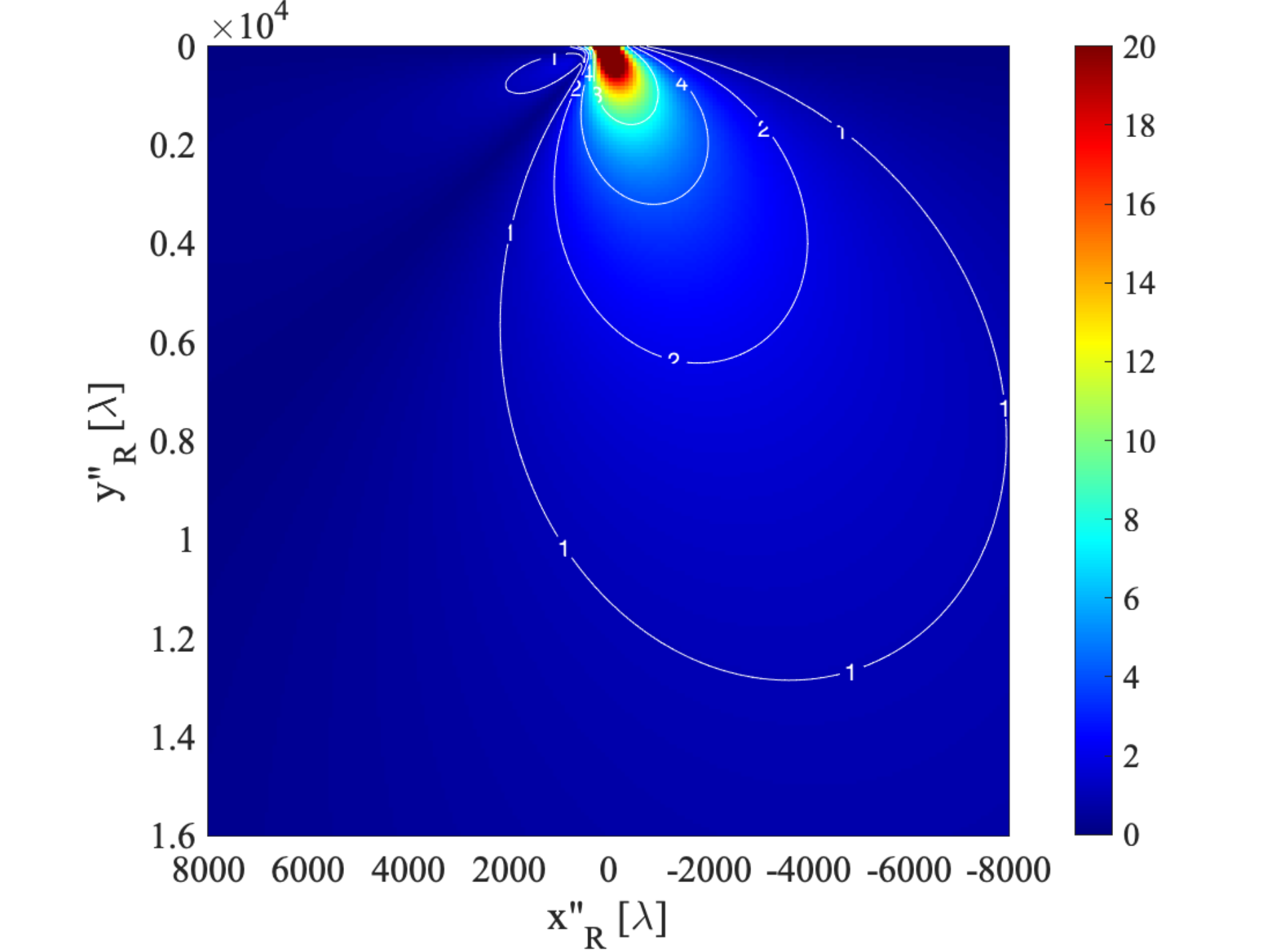} }
		\subfigure[$\varphi = \pi/2$] {\includegraphics [width=.41\linewidth, trim=10 210 30 220, clip]{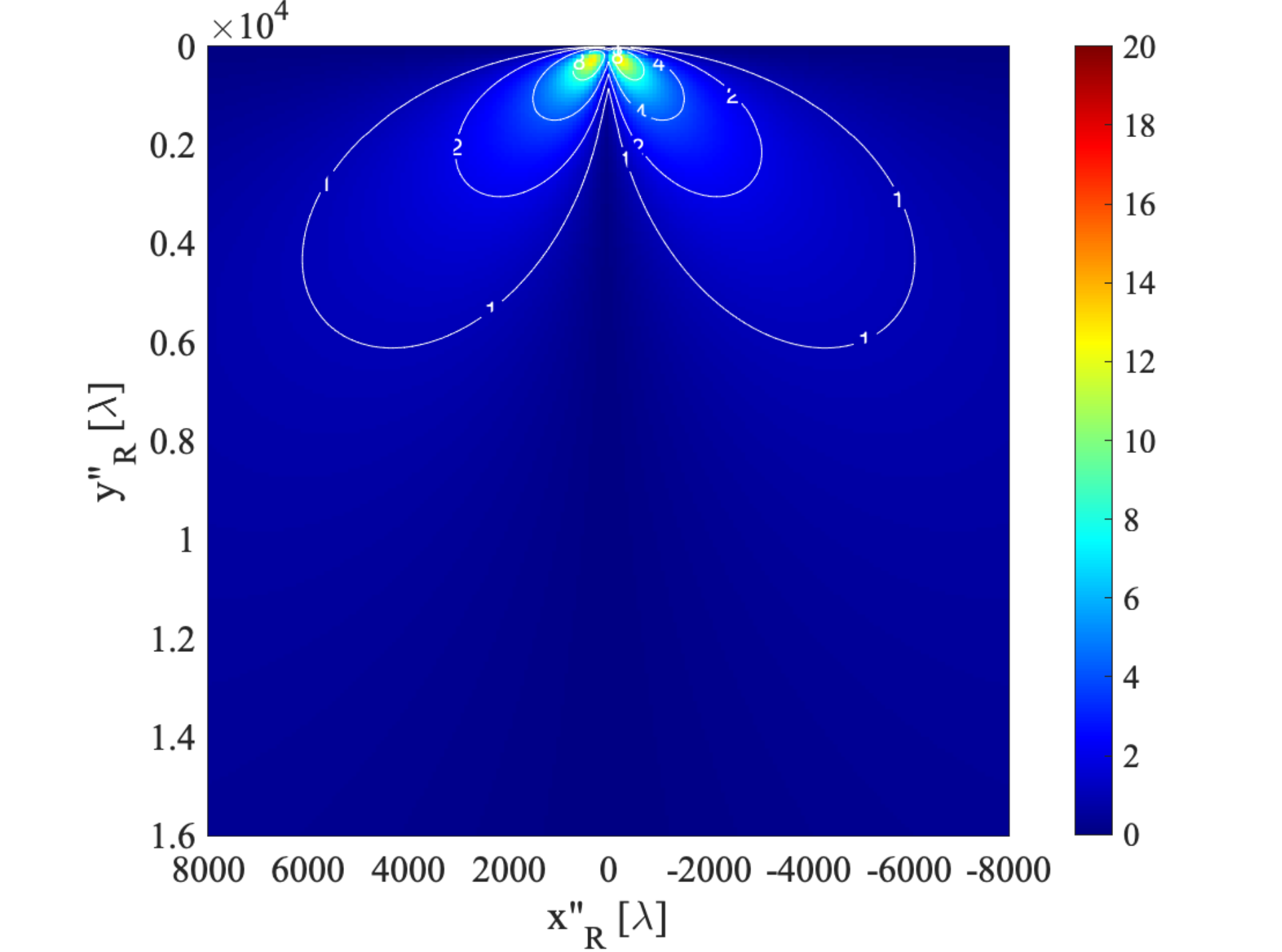} }\hspace{0.8cm}
		\subfigure[$\varphi = \varphi_h({\mathrm{o}} )$]  {\includegraphics[width=.41\linewidth, trim= 10 210 30 220, clip]{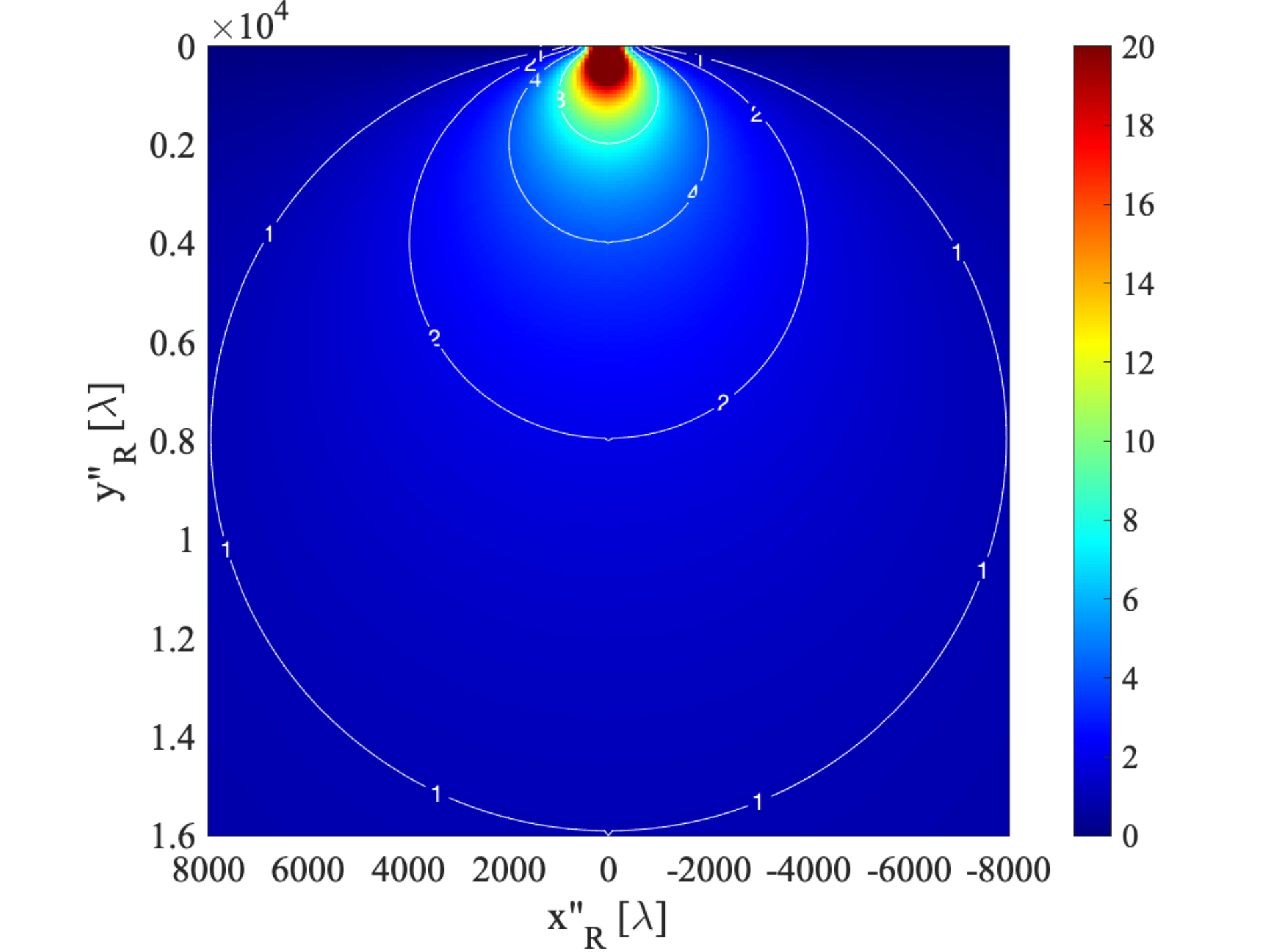} }	
		\caption{The K numbers achieved by $\mathcal L_r$ with different orientation angles on the $+\mathrm{y}''$-half ground plane in the horizontal LSAA placement scenario, with  $L = 400 \lambda$,  $\rho = 20 \lambda$, and  $z''_{S,h} = 200 \lambda$. Contour lines at $K = 1$, $2$, and $4$ are shown.}
		\label{fig:DoF_hTx}
	\end{figure*}
	
This scenario is shown in  Fig.~\ref{fig8}(b). Following Section~\ref{sec:4a}, an  $\mathrm{o}'$-origin coordinate system and an $\mathrm{o}$-origin receiving coordinate system are defined. The $\mathrm{x}''$-axis,  $\mathrm{z}'$-axis, and $\mathrm{z}$-axis are all in parallel. Letting $r_2   = \sqrt{ (y''_R)^2 + ( z''_{T,h})^2 }$,  the polar angle is given by $\theta = \arctan^* \big( { r_2  }/{x''_R}\big)$.  The unit directional vector of the $\mathrm{x}$-axis is given by $\hat{\mathbf{e}}_{x} =  \big(0, {y''_R}/{r_2}, - {z''_{T,h}}/{r_2} \big)$.  As shown in the {left-view} diagram in Fig.~\ref{fig8}(b), we define the angle between the $\mathrm{x}$-axis and the $\mathrm{y}''$-axis, which is also the angle between the ground plane and the  $\mathrm{o}$-$\mathcal L_s$ plane, to be  $\psi ({\mathrm{o}})  \triangleq  \arccos ( \langle \hat{\mathbf{e}}_{x}, \hat{\mathbf{e}}_{y''}\rangle ) = \arccos \big( {y''_R}/{r_2} \big)$. 
	
The projections of $\mathcal L_r$ in the three orthogonal receiving directions are given by  $L^h_{R,z} = L_r |\cos\varphi|$,  $L^h_{R,x}  = L_r \sin\varphi \cos \psi({\mathrm{o}})  $, and  $L^h_{R,y}  = L_r \sin\varphi \sin \psi({\mathrm{o}}) $, respectively. Since all receiving directions, especially $ \hat{\mathbf{e}}_{z}$, may contribute, better K number performance can be expected compared to the vertical placement scenario.  When $y''_R$ is very small,  $\psi ({\mathrm{o}})$ will be close to $\pi/2$, leading to $L^h_{R,x}  \approx 0$ and a large projection $L^h_{R,y}  \approx 2\rho \sin\varphi $. When $y''_R  \gg z''_{S,h}$,  $\psi({\mathrm{o}}) $ becomes very small, causing  $L^h_{R,x}  \approx 2\rho \sin\varphi $ and $L^h_{R,y} \approx 0$. Thus, as a trad-off between increasing $r$ and $L^h_{R,x} $ and decreasing $| \theta -  \pi/2 |$,  the best K number performance is expected to appear at some small positive value of $y''_R$.  
	
Nevertheless, $\mathcal L_r$ can hardly fall inside the spatial multiplexing region in the $\hat{\mathbf{e}}_{y}$ receiving direction. Following the rule of thumb, we focus on the contributions from the rest two receiving directions. Moreover, from Fig.~\ref{fig5}(a) and Fig.~\ref{fig5}(b), we know that owing to the elevated placement, the changes in the spatial bandwidth in $\hat{\mathbf{e}}_{z}$ and $\hat{\mathbf{e}}_{x}$ directions can also be ignored. Accordingly, in case $\mathcal L_r$ can be rotated on demand, a simple orientation control solution we suggest is as follows: 
	\begin{equation}
		\varphi_h({\mathrm{o}} ) = \arctan^* \Big(\mathrm{sign}^*\big(- x''_{R}\big)  \frac{\cos\psi({\mathrm{o}})    w_{x}(0;\mathbf{\Omega}) }{ w_{z}(0;\mathbf{\Omega}) } \Big)
	\end{equation}
where the $\mathrm{sign}^*(x)$ operator returns $1$ if $x >0$, $0$ if $x =0$, and $-1$ if $x< 0$. It ensures that $\varphi_h({\mathrm{o}} )  > \pi/2$ when $ x''_{R} > 0$,  $\varphi_h({\mathrm{o}} ) < \pi/2$ when $ x''_{R} < 0$, and 
	\begin{equation*}
		\frac{\sin\varphi_h({\mathrm{o}} )  }{ | \cos \varphi_h({\mathrm{o}} ) | } =  \frac{\cos\psi({\mathrm{o}})  w_{x}(0;\mathbf{\Omega}) }{ w_{z}(0;\mathbf{\Omega}) }
	\end{equation*}
for both cases; and additionally, $\varphi_h({\mathrm{o}} ) = 0$ when $ x''_{R}  = 0$. We note that only if the following two conditions are met, can $\varphi_h({\mathrm{o}} )$ be considered optimal in the practical sense: First, $\mathcal L_r$ is located in the constant bandwidth spatial multiplexing regions for both $\hat{\mathbf{e}}_{z}$ and  $\hat{\mathbf{e}}_{x}$ directions, whose boundaries should be determined using $\rho = L^h_{R,z}/2$  and $\rho = L^h_{R,x}/2$, respectively. This condition is to ensure that the local spatial bandwidth seen along $\mathcal L_r$ can be considered constant and approximated using $w_{z}(0;\mathbf{\Omega}) $ and $w_{x}(0;\mathbf{\Omega})$. Second, $r |\cos\theta| > \frac{L}{2}$, which ensures that the maximum and the minimum spatial frequency components seen in both $\hat{\mathbf{e}}_{z}$ and  $\hat{\mathbf{e}}_{x}$ directions are caused by the same ends of $\mathcal L_s$, as can be seen from \eqref{eq:spatial_bandwidth_wz} and \eqref{eq:spatial_bandwidth_wx}, and thus the following equation holds: $$  w_{\hatbf v}(0;\mathbf{\Omega}) = |\cos\varphi| w_{z}(0;\mathbf{\Omega}) +  \sin\varphi \cos \psi({\mathrm{o}}) w_{x}(0;\mathbf{\Omega}). $$ 
Accordingly, the orientation angle $\varphi_h({\mathrm{o}} )$ maximizes  $w_{\hatbf v}(0;\mathbf{\Omega})$ and hence the K number. Note that the $\hat{\mathbf{e}}_{y}$ direction is considered to make no contribution. Therefore, although optimality cannot be guaranteed in all situations, $\varphi_h({\mathrm{o}} )$ can be expected to ensure good performance.

Fig.~\ref{fig:DoF_hTx}  presents the color maps of the K number obtained numerically following \eqref{eq:spatial_DoF_RCS} (a cutoff at $20$ is applied), achieved by  $\mathcal L_r$ at different locations  on the $+\mathrm{y}''$-half ground plane,  with three fixed orientation angles $\varphi = 0$, $\pi/4$ and $\pi/2$ and the location-dependent one $ \varphi  = \varphi_h({\mathrm{o}} )$.  We remind that the area shown in Fig.~\ref{fig:DoF_hTx} is much larger that in Fig.~\ref{fig:DoF_vTx}. The similarity in the shapes of the regions bounded by the contour lines in  Fig.~\ref{fig:DoF_hTx}(a) and Fig.~\ref{fig:DoF_hTx}(c) and the spatial multiplexing regions in Fig.~\ref{fig4}(a) and Fig.~\ref{fig4}(b) can be clearly seen, and the expected offsets in the positions seeing the largest K number from the $\mathrm{x}''$-axis, as discussed above, are also verified. As shown by Fig.~\ref{fig:DoF_hTx}(d), with the simple orientation control $ \varphi  = \varphi_h({\mathrm{o}} )$, a larger K number coverage is achieved, whose shape is roughly a disk. Compared with Fig.~\ref{fig:DoF_hTx}(a), its widening in the $\mathrm{x}''$-axis direction is owing to the contribution from the $\hat{\mathbf{e}}_{x}$ receiving direction. As in the vertical placement scenario, significant performance degradation may occur at some position if the orientation of  $\mathcal L_r$ is fixed. 
	
\section{Conclusions} 
In this paper, a unified spatial bandwidth viewpoint has been adopted to review the theoretical basis of the achievable spatial DoF in LOS channels in LSAA-based communications. With the aim to fill some of the knowledge gaps regarding the impact of the 3D linear array assembly geometry, a receiving coordinate system and parameterization strategy have been proposed to reduce the dimensionality of a complex problem into three elementary ones in three orthogonal receiving directions.  An analytical framework based on spatial bandwidth analysis has been developed, under which simple and accurate closed-form approximation formulas for the achievable spatial DoF have been derived and spatial multiplexing regions have been defined. With these simple closed-form formulas and the visualized spatial multiplexing regions, additional insights on the underlying mechanism of spatial DoF can be drawn, the influence of the array orientation can be quantified, and its behavior under different geometric conditions can be more easily predicted. 
	
The obtained results show that, surprisingly, the constant spatial bandwidth approximation is generally valid for the $\hatbf{e}_x$ and $\hatbf{e}_z$ directions. Although the same claim cannot be made for the $\hatbf{e}_y$ direction, its contribution to the achievable spatial DoF is minor as compared with the other two and hence can be neglected in many practical conditions. With these findings and the closed-form approximation formulas, large-scale system-level simulation research can be significantly accelerated and more extensive analytical research can also be facilitated. In addition,  unexpectedly, the $\hatbf{e}_x$ direction can contribute more than the $\hatbf{e}_z$ direction in some locations within the desired coverage area.  Furthermore, based on these findings, in the special setups described in the case study, a close-to-optimal orientation of the linear receiving array is given in simple and closed forms and good achievable spatial DoF coverage performances are shown, demonstrating the usefulness and benefits of the proposed analytical framework. 

Finally, we would like to remark that the numerical studies in this paper have been focused on the influence of position and rotation owing to space limitations. The impact of the array size can also be studied using the same framework. 
	
\bibliographystyle{IEEEtran}
\bibliography{ref}

\end{document}